\documentclass[twocolumn ]{aastex63}

\usepackage{amsmath}

\submitjournal{ApJ}

\shorttitle{Radio/ultraviolet/optical sizes of  massive galaxies}
\shortauthors{Jim\'enez-Andrade et al.}

\begin{document}

\title{The VLA Frontier Field Survey:\\ a comparison of the radio and UV/optical  size of  $0.3 \lesssim z \lesssim 3$ star-forming galaxies}

\correspondingauthor{Jim\'enez-Andrade, E.F.}
\email{ejimenez@nrao.edu}

\author[0000-0002-2640-5917]{E.~F. Jim\'enez-Andrade}
\affiliation{National Radio Astronomy Observatory, 520 Edgemont Road, Charlottesville, VA 22903, USA}

\author[0000-0001-7089-7325]{E.~J. Murphy}
\affiliation{National Radio Astronomy Observatory, 520 Edgemont Road, Charlottesville, VA 22903, USA}  

\author[0000-0001-6864-5057]{I. Heywood}
\affiliation{Astrophysics, Department of Physics, University of Oxford, Keble Road, Oxford, OX1 3RH, UK}  
\affiliation{Department of Physics and Electronics, Rhodes University, PO Box 94, Makhanda, 6140, South Africa}
\affiliation{South African Radio Astronomy Observatory, 2 Fir Street, Observatory, 7925, South Africa}  

\author[0000-0003-3037-257X]{I. Smail}
\affiliation{Centre for Extragalactic Astronomy, Department of Physics, Durham University, South Road, Durham DH1 3LE, UK}

\author{K. Penner}
\affiliation{National Radio Astronomy Observatory, 520 Edgemont Road, Charlottesville, VA 22903, USA}  

\author[0000-0003-3168-5922]{E. Momjian}
\affiliation{National Radio Astronomy Observatory, P.O Box O, Socorro, NM 87801, USA}  

\author[0000-0001-5414-5131]{M. Dickinson}
\affiliation{National Optical Astronomy Observatories, 950 N Cherry Avenue, Tucson, AZ 85719, USA}  

\author[0000-0003-3498-2973]{L. Armus}
\affiliation{Infrared Processing and Analysis Center, MC 314-6, 1200 E. California Boulevard, Pasadena, CA 91125, USA}  

\author{T.~J.~W. Lazio}
\affiliation{Jet Propulsion Laboratory, California Institute of Technology, 4800 Oak Grove Drive, Pasadena, CA 91109, USA}  



\begin{abstract}
To investigate the growth history of galaxies, we measure the rest-frame radio, ultraviolet (UV), and optical sizes of 98 radio-selected, star-forming galaxies (SFGs) distributed over $0.3 \lesssim z \lesssim 3$ and median stellar mass of $\log(M_\star/ \rm M_\odot)\approx10.4$. We  compare the size of galaxy stellar disks, traced by rest-frame optical emission, relative to the overall extent of star formation activity that is traced by radio continuum emission.  Galaxies in our sample are identified in three   {\it Hubble} Frontier Fields:  MACS\,J0416.1$-$2403, MACS\,J0717.5+3745, and MACS\,J1149.5+2223. 
Radio continuum sizes are derived from  3\,GHz and 6\,GHz radio images ($\lesssim 0\farcs6$\, resolution, $\approx0.9\, \rm \mu Jy\, beam^{-1}$ noise level) from the {\it Karl G. Jansky} Very Large Array.   
Rest-frame UV and optical sizes are derived using observations from the {\it Hubble Space Telescope} and the ACS and WFC3   instruments. 
We find no clear dependence between the 3GHz radio size and stellar mass of SFGs, which contrasts with the positive correlation between the UV/optical size and stellar mass of galaxies. 
Focusing on SFGs with $\log(M_\star/\rm M_\odot)>10$, we find that the radio/UV/optical emission tends to be more compact in galaxies with high star-formation rates ($\rm SFR\gtrsim 100\,M_\odot\,yr^{-1}$), { suggesting that a central, compact starburst (and/or an Active Galactic Nucleus) resides in the most luminous galaxies of our sample. }
We also find that
the physical radio/UV/optical size of { radio-selected}  SFGs with $\log(M_\star/\rm M_\odot)>10$ increases by a factor of $1.5-2$ from $z\approx 3$ to $z\approx0.3$, yet   the  radio emission remains  two-to-three times more compact than that from the UV/optical. These findings indicate 
that these massive, {radio-selected} SFGs at  $0.3 \lesssim z \lesssim 3$ tend to harbor centrally enhanced star formation activity relative to their outer-disks.
\end{abstract}



\section{Introduction} \label{sec:intro}

The stellar mass buildup in galaxies is thought to be regulated by secular and violent/stochastic evolutionary channels \citep[e.g.,][]{steinhardtspeagle14, madau14}, including steady  inflows of gas from the cosmic web \citep[e.g.,][]{dekel09}, galactic winds \citep[e.g.,][]{veilleux05}, and galaxy-galaxy interactions or mergers \citep[e.g.,][]{hopkins06}. Because these processes leave strong signatures on the galaxy structure \citep{conselice14, tacchella16, habouzit19},  measuring the size of high-redshift galaxies has become critical to investigate  the growth history of present-day galaxy populations. 
 
The size evolution of galaxies has been primarily investigated using optical and near-infrared  imaging with the {\it Hubble Space Telescope} ({\it HST}), which has most recently been obtained with the    Advanced Camera for Surveys (ACS)  and  Wide Field Camera 3 (WFC3), respectively.  The consensus is that the size of the stellar component of  galaxies increases with redshift,  stellar mass, and/or luminosity \citep[e.g.,][]{bouwens04, ferguson04, franx08, grazian12,  mosleh12, huang13, morishita14, vanderwel14, shibuya15, ribeiro16}. By  linking the size of galaxies with their level of star formation, it is believed that quiescent galaxies grow across cosmic time at a higher rate than star-forming galaxies \citep[SFGs;][]{morishita14, vanderwel14}, likely because gas-poor minor mergers efficiently enlarge the size of massive, passive galaxies \citep{hiltz12, carollo13, morishita14, vandokkum15}.  At a fixed stellar mass, quiescent and starburst galaxies tend to be more compact than most SFGs on the main sequence  \citep[MS;][]{wuyts11, gu20}. This finding suggests that strong inflows of gas that enhance the galaxy's central density can trigger a starburst, while compact quenched galaxies have a high stellar density leftover from such a burst (\citealt{tacchella15, faisst17, wang19, wu20}, but see \citealt{abramson18}). Since  the slope of the size -- stellar mass relation of quiescent and  SFGs  appears to be invariant with redshift \citep{vanderwel14}, it is expected that the assembly mechanisms of these  galaxy's populations act similarly at all cosmic epochs. 

Observations of the rest-frame optical continuum --tracing the galaxy's stellar component--  are insufficient to investigate the spatial distribution of star formation, which is actively contributing to the stellar mass buildup in galaxies.  H$\alpha$ and ultraviolet (UV) continuum observations can be used to highlight the  population of young, massive stars in galaxies \citep[e.g.,][]{salim07, dudzeviciute20}.  However, especially at high redshifts, massive SFGs harbor  high concentrations of interstellar dust \citep[e.g.,][]{calura17, dudzeviciute20} that obscure the H$\alpha$/UV emission \citep{buat12, nelson16, chen20}.   One dust-unbiased, but indirect, probe of star formation can be obtained by mapping the {predominantly non-thermal} radio continuum emission of galaxies at centimeter wavelengths \citep[e.g.,][]{condon92, bell03, garn09, murphy11}.   This star-formation rate (SFR) indicator has been calibrated  using the tight, yet empirical, far-infrared (FIR) -- radio correlation \citep[][]{helou85, yun01, murphy06, murphy06b, murphy09,  murphy12, sargent10, magnelli15, delhaize17, gim19, algera20}. 
The physical interpretation of this relation is that FIR emission arises from the absorption and re-radiation of UV and optical photons that heat dust grains surrounding massive star-forming regions. 
These same massive stars end their lives as core-collapse supernovae, whose remnants (SNRs) accelerate cosmic ray electrons (CREs) throughout the {magnetized interstellar media (ISM)} of galaxies,  producing diffuse non-thermal synchrotron emission at radio frequencies \citep[e.g.,][]{condon92, helou93}. As a result, there is a close (although complex) correlation between the spatial  distribution of young stars and the observed synchrotron radio  emission of galaxies \citep[e.g.,][]{lequeux71,  heesen14}. Due to the propagation of CRE across galactic disks \citep[from the SNRs to their current location of emission; e.g.,][]{murphy06, murphy08}, the radio synchrotron map of galaxies can generally be described by a smoothed version of the source distribution of CREs.  Thus,   galaxy-averaged, radio size measurements are a proxy for the overall extent of massive star formation in galaxies \citep[see][]{heesen14}.

Since  large-scale extragalactic surveys at high-angular resolution can be efficiently produced with the {\it Karl G. Jansky} Very Large Array (VLA), radio continuum surveys have been paramount to identify large samples of SFGs to  trace the dust-unbiased production of stars at high redshifts \citep[e.g.,][]{richards00, schinnerer10,owen18, smolcic17}. Recent studies have shown that the radio continuum size of galaxies decreases with increasing redshift \citep{bondi18, lindroos18, jimenezandrade19}, and that the radio size of starbursts tends to be more compact than that of typical galaxies on the main sequence \citep[][]{murphy13, jimenezandrade19, thomson19}. These observations have also suggested that star formation is centrally concentrated in galaxies out to $z\approx2$, as the radio emission of galaxies remains $\sim2$ times more compact than the optical light  \citep{murphy17, bondi18, lindroos18, owen18, jimenezandrade19, thomson19} that traces most of the stellar mass in galaxies. High-resolution, dust continuum observations taken with the  Atacama Large Millimeter/submillimeter Array (ALMA) have revealed similar trends, i.e., that the stellar morphologies of $z>1$ galaxies appear significantly more extended than the dust continuum emission tracing star formation \citep[e.g.,][]{simpson15, hodge16, gullberg19}. 
  
In summary,  while {\it HST} observations  permit detailed characterization of  the growth history of the stellar component in galaxies, recent  VLA observations are reaching sufficient angular resolution and sensitivity to investigate where new stars form in galaxies across cosmic time. The  VLA and {\it HST}  enable  combined analyses of radio and UV/optical imaging to  \emph{simultaneously} measure the  multi-wavelength size  of high-redshift galaxies, allowing us to explore whether the radio size follows similar  evolutionary trends as  the UV/optical one.  

In this era of multi-wavelength astronomy, deep panchromatic observations  are available toward extragalactic fields.  This is the case of the {\it Hubble} Frontier Fields (HFF) project \citep{lotz17} that provides ACS and WFC3 imaging toward  six massive galaxy clusters  and their parallel fields. A  wealth of ancillary data is also available towards the HFF, including observations from {\it Herschel} \citep{rawle16}, ALMA \citep{gonzalezlopez17, laporte17}, {\it Spitzer} \citep{lotz17}, {\it Chandra} observatories \citep{vanweeren17}   as well as ground-based telescopes \citep[see][and references therein]{shipley18}.  These observations have been recently complemented by the VLA Frontier Field project (Heywood et al., {{ in press}}), which delivers sub-arcsec radio continuum imaging at 3 and 6\,GHz toward three  HFF clusters.  The sensitivity of the VLA Frontier Field images of $\approx0.9\,\mu \rm Jy \, beam^{-1}$ allows us to probe the faint-end of the  radio source population, which is primarily powered by star formation processes and not Active Galactic Nuclei  \citep[AGN; ][]{smolcic17b}. 

In this paper, we combine {\it HST} and VLA imaging from the HFF project to measure the radio size of  galaxies  relative to those from  UV/optical emission. We  explore the dependence of the radio and UV/optical  size on the redshift, stellar mass, and SFR. 
We use this knowledge to better understand the mechanisms driving  the growth of galaxies over $0.3\lesssim z \lesssim 3$, which is the  cosmic epoch during which most of the stellar mass observed today was assembled \citep[e.g.,  ][]{behroozi13, madau14}. 

This paper is organized as follows.  The data set used in this study is described in \S  \ref{sec:data}. In \S \ref{sec:methods},  we describe the sample selection  and derive radio/UV/optical size of 98 galaxies in the sample. In \S \ref{sec:results}, we present the results and discuss  the implications of this work on the current picture of galaxy evolution. We use the AB magnitude system and adopt a flat $\Lambda$CDM cosmology with $h_0 = 0.7$, $\Omega_M = 0.3$, and $\Omega_\Lambda = 0.7$, to be consistent  with the cosmological parameters used to construct the HFF photometric catalogs used here.

\section{Data}
\label{sec:data}

 \subsection{VLA images}
We use the radio continuum maps at 3\,GHz (S-band) and 6\,GHz (C-band) produced by Heywood et al.\;({{ in press}}) as part of the VLA Frontier Fields survey (PI: E. Murphy; VLA/14A-012, 15A-282, 16B-319). These images were obtained by combining VLA observations with the A and C configuration,   achieving high spatial resolution without missing diffuse and extended  emission of radio sources. The VLA images at 3\,GHz of the MACS\,J0416-2403, MACS\,J0717+3745,  and MACS\,J1149+2223 fields  (hereafter MACS\,J0416, MACS\,J0717,  and MACS\,J1149, respectively)  have a native resolution {measured at the full width half maximum (FWHM)}  of $0\farcs94\times 0\farcs51$, $0\farcs73\times0\farcs61$, and $0\farcs51\times0\farcs48$.  At 6\,GHz, these maps have a { FWHM} resolution of $0\farcs53\times 0\farcs30$, $0\farcs33\times0\farcs27$, $0\farcs28\times0\farcs27$, respectively. The single-pointing VLA observations at 3\,GHz (6\,GHz) are centered at their respective cluster field and extend out to a radius of 12\arcmin~(6\arcmin) at the 30\% primary beam level. The  large extent of the 3\,GHz maps suffices to retrieve radio emission from galaxies in the HFF clusters and their associated parallel fields, while the smaller radius of the  maps at 6\,GHz is enough to cover the  HFF core fields and only a small fraction ($\sim 25\%$) of their parallel fields (Heywood et al. {{ in press}}). A $\mu $Jy-level sensitivity was achieved in all the maps  at 3 and 6\,GHz, with a typical $1\sigma$ noise of $\approx 0.9\,\rm \mu Jy \,beam^{-1} $. We refer the reader to Heywood et al. ({in press}) for further details on the VLA data reduction, image production, and source extraction. 

\begin{figure*}
	\begin{centering}
		\includegraphics[width=18cm]{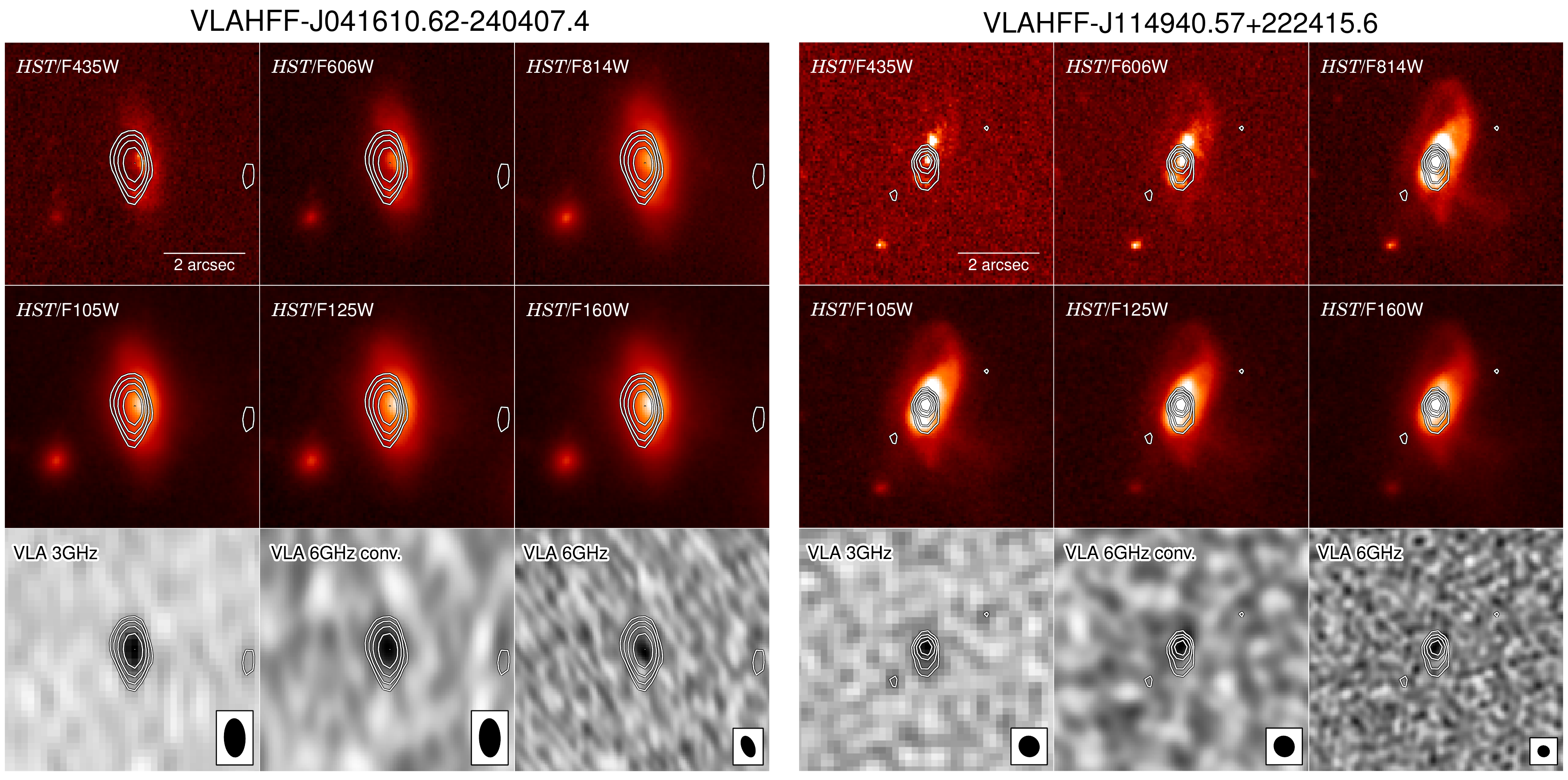}
		\caption{The  multi-wavelength view of the morphology of galaxies in the HFF.  Two examples are presented:  VLAHFF-J041610.62-240407.4 at $z=0.41$ and VLAHFF-J114940.57+222415.6 at $z=0.76$. The two upper rows show stamps from the six {\it HST} bands used in this study, while the bottom row presents the stamps from the VLA map at 3 and 6\,GHz (including the original  and the convolved image to the resolution of the 3\,GHz  map). The size of the synthesized  beam is shown at the bottom-right corner of the stamps. The contour levels start at $3\sigma\simeq 3\rm \mu Jy\,beam^{-1}$ and are spaced on a logarithmic scale.  The stamps for all the galaxies in the sample are available at \url{https://science.nrao.edu/science/surveys/vla-ff/data}.}
		\label{fig:stamps}
	\end{centering}
\end{figure*}

\subsection{{\it HST} data}

We use {\it HST} observations taken with the ACS and WFC3 instruments as part of the HFF program \citep[][Koekemoer et al.\;submitted]{lotz17}. We employ the images from the wide-band filters ACS/{\it F}435W,  ACS/{\it F}606W, ACS/{\it F}814W, WFC3/{\it F}105W, WFC3/{\it F}125W, and WFC3/{\it F}160W.   Combined, these images continuously cover the observed wavelength range of 0.35$-$1.7$\,\mu$m, i.e., the optical-to-near-IR regime (see Figure\,\ref{fig:stamps}). For galaxies over the redshift range $1 \lesssim z \lesssim 2$  we can simultaneously observe the rest-frame UV (0.15$-$0.3$\,\mu$m) and optical  (0.45$-$0.8$\,\mu$m) continuum emission. 
For galaxies at $z\lesssim 1$, we can observe their rest-frame optical emission through the ACS {\it HST} filters,  while the  WFC3 {\it HST} filters trace the rest-frame UV emission of $z\gtrsim 2$ galaxies.

We employ the final mosaics (with 60 mas pixel scale) of MACS\,J0416-2403, MACS\,J0717+3745,  and MACS\,J1149+2223 (cluster and parallel fields) produced by the HFF DeepSpace team\footnote{\url{http://cosmos.phy.tufts.edu/~danilo/HFF/Download.html}} \citep[][]{shipley18}, which, combined,  encompass a sky area of $\approx68\,\rm arcmin^2$.  These are stacked and drizzled images (version v1.0) from the HFF program \citep[][Koekemoer et al.\;submitted.]{lotz17}, in which  major artifacts and cosmic rays were removed. We  also download  the segmentation maps and PSFs, required in this work,  that have been produced and released to the public by the  HFF DeepSpace team. The PSFs have been empirically derived by stacking non-saturated stars per each of the HFF fields and {\it HST} filters \citep{shipley18}.  A 2D Gaussian fit to the PSFs indicates an average ${\rm FWHM}\approx 0\farcs11$ and 0\farcs18 resolution for the ACS and WFC3 images.

\subsection{The VLA Frontier Fields  catalogs}

Our galaxy sample is drawn from the VLA Frontier Fields catalog  (Heywood et al.\;{in press}). This compilation contains information for 1966  and 257 compact radio sources detected  at 3\,GHz and 6\,GHz, respectively, across the three HFF cluster and parallel fields explored here. A minimum signal-to-noise ratio ($S/N$) threshold of 5 was adopted to detect the radio sources with the {\sc PyBDSF} code \citep{Mohan15}. Flux density and FWHM estimates of radio sources (before and after deconvolution)  are also reported in the catalog. 

The VLA Frontier Fields catalog also reports spectroscopic/photometric redshifts and  stellar masses of radio sources with a counterpart in the  HFF-DeepSpace Photometric Catalogs \citep{shipley18}. 113 radio-selected galaxies with available redshift and stellar mass estimates are identified.  The low fraction of 3\,GHz-detected sources with a counterpart in the HFF-DeepSpace  Catalogs is a result of the  large sky coverage of the VLA 3\,GHz radio maps, which is $\sim 14$ times  that of the {\it Hubble} Frontier Fields themselves.
Photometric redshifts were derived by \citet{shipley18} via {  spectral energy distribution (SED)} fitting with the {\sc eazy} code \citep{brammer08}.  The photometric redshift used here corresponds to the peak of the redshift distribution. These photometric estimates are found to be in good agreement with available spectroscopic redshifts, as \citet{shipley18} finds an average dispersion  of $\sigma=0.034$ and a gross failure rate of $\lesssim10\%$. Using such redshift estimates, Heywood et al.\;({  in press})  estimated  the lensing magnification factors for these radio sources using all available lensing models from the Frontier Fields team\footnote{\url{https://archive.stsci.edu/prepds/frontier/lensmodels/} } \citep[e.g.,][]{juazac12, juazac14, johnson14}. The median magnification factor $(\mu)$, recorded in the VLA Frontier Fields catalog,  is used throughout this paper to derive the intrinsic properties of our galaxy sample. Photometric redshifts, or spectroscopic estimates when available, were also used  by  \citet{shipley18}  to derive stellar masses with the  {\sc fast} code \citep{kriek09}. To this end, the \citet{bruzual03} stellar population synthesis model library and the \citet{calzetti00} dust attenuation law  was employed. A \citet{chabrier03} IMF, solar metallicity, and exponentially declining star formation histories were adopted as well. These stellar mass  estimates ($M_\star^{\rm FAST}$)  are here corrected  for lensing magnification as follows:  $M_\star=M_\star^{\rm FAST}/\mu$, where $\mu$ is the median magnification factor reported by Heywood et al.\;({  in press}).

\section{Analysis}
\label{sec:methods}
\subsection{The sample}
\label{subsec:sample}
We select the 113  galaxies in the VLA Frontier Fields 3\,GHz catalog  with available   redshift and  stellar mass estimates from  the HFF DeepSpace catalogs. 31 of these are also detected at 6\,GHz. A total of 14 (7), 38 (15) and 32 (7) sources are identified in the MACS\,J0416, MACS\,J0717 and MACS\,J1149 cluster (parallel) fields, respectively.  The smaller numbers of sources in the parallel fields is a result of the cluster overdensity, as well as the  reduced radio sensitivity and negligible lensing magnification at the positions of the parallel fields.

\subsubsection{Selecting field galaxies in the HFF}
We remove cluster galaxies from the sample to account for the potential biases linked to the environmental dependence of galaxy evolution. This is done by selecting sources that are inside the projected cluster virial radius, and that have line-of-sight velocities falling within the velocity dispersion of galaxies in the clusters. In this case, the virial radius of these complexes \citep[$\sim$2\,Mpc/5\arcmin; e.g., ][]{balestra16} is larger than the extent of the {\it HST} mosaics of the HFF clusters ($\sim 2\arcmin$ radius). As a result, the selection of cluster galaxies only relies on their line-of-sight velocity ($v_{\rm gal}={\rm c} z_{\rm gal}$). Accordingly, we recognize cluster galaxies as those satisfying the condition $|z_{\rm gal} -z_{\rm cluster}|<3\sigma_{z}$, where  $z_{\rm cluster}$  is the redshift  of the HFF clusters, $\sigma_{z}$ the associated redshift  dispersion, and  $z_{\rm gal}$   the redshift of the galaxy of interest. The values of  $z_{\rm cluster}$ $(\sigma_{z})$ are 0.3972 (0.0033), 0.5450 (0.0036), and 0.5422 (0.0048) for MACS\,J0416 \citep{balestra16}, MACS\,J0717  \citep{richard14},  and MACS\,J1149 \citep{grillo16}, respectively. This analysis identifies 15 cluster galaxies out of the initial 113 galaxies, leading to a sample of 98 field galaxies  with available redshift and stellar mass estimates. Although the focus of this study is the nature of field galaxies,  in Table \ref{table_sllsizes}, we also report the radio/UV/optical sizes of those 15 cluster galaxies.

\subsubsection{Magnification factor, redshift, and stellar mass distribution}
\begin{figure*}
	\begin{centering}
		\includegraphics[width=18.3cm]{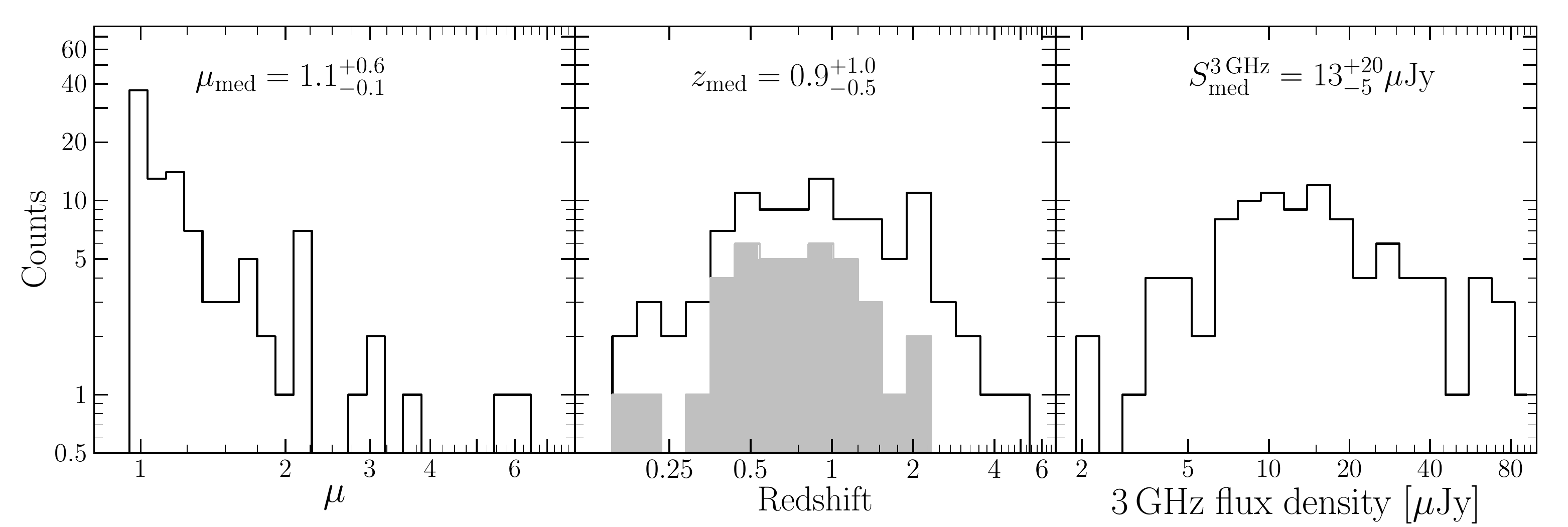}	
		\caption{Distribution of the magnification factor ({\it left panel}), redshift ({\it middle panel}), and 3\,GHz flux density ({\it right panel}) of 98 radio-selected, field galaxies in the HFF fields,  for which information about their stellar masses and UV/optical photometry are available in the HFF DeepSpace catalogs.  The median of the distributions is shown at the top, the lower/upper limits correspond to the 16th/84th percentile of the distribution. The grey filled histogram shows the distribution of 40 galaxies with available spectroscopic redshift, whose median is $0.7^{+0.4}_{-0.3}$. The 3\,GHz flux density has been corrected by lensing magnification.}
		\label{fig:mu_distribution}
	\end{centering}
\end{figure*}

The median magnification factor of the 98 field galaxies in our sample is $\mu=1.1^{+0.6}_{-0.1}$, where the upper/lower limit correspond to the 84th/16th percentile of the distribution. 
Only a small fraction ($\approx13\%$)  are moderately lensed with magnification factors between 2 and 6 (Figure\,\ref{fig:mu_distribution}). Because morphological parameters of lensed galaxies with $\mu<30$ can be derived without introducing strong biases \citep[][]{florian16}, we also include moderately lensed galaxies in our sample.  We  verified that none of the results presented here significantly change if sources with $\mu \geq 2$ are excluded from the analysis.


\begin{figure*}	
   \includegraphics[width=9.2cm]{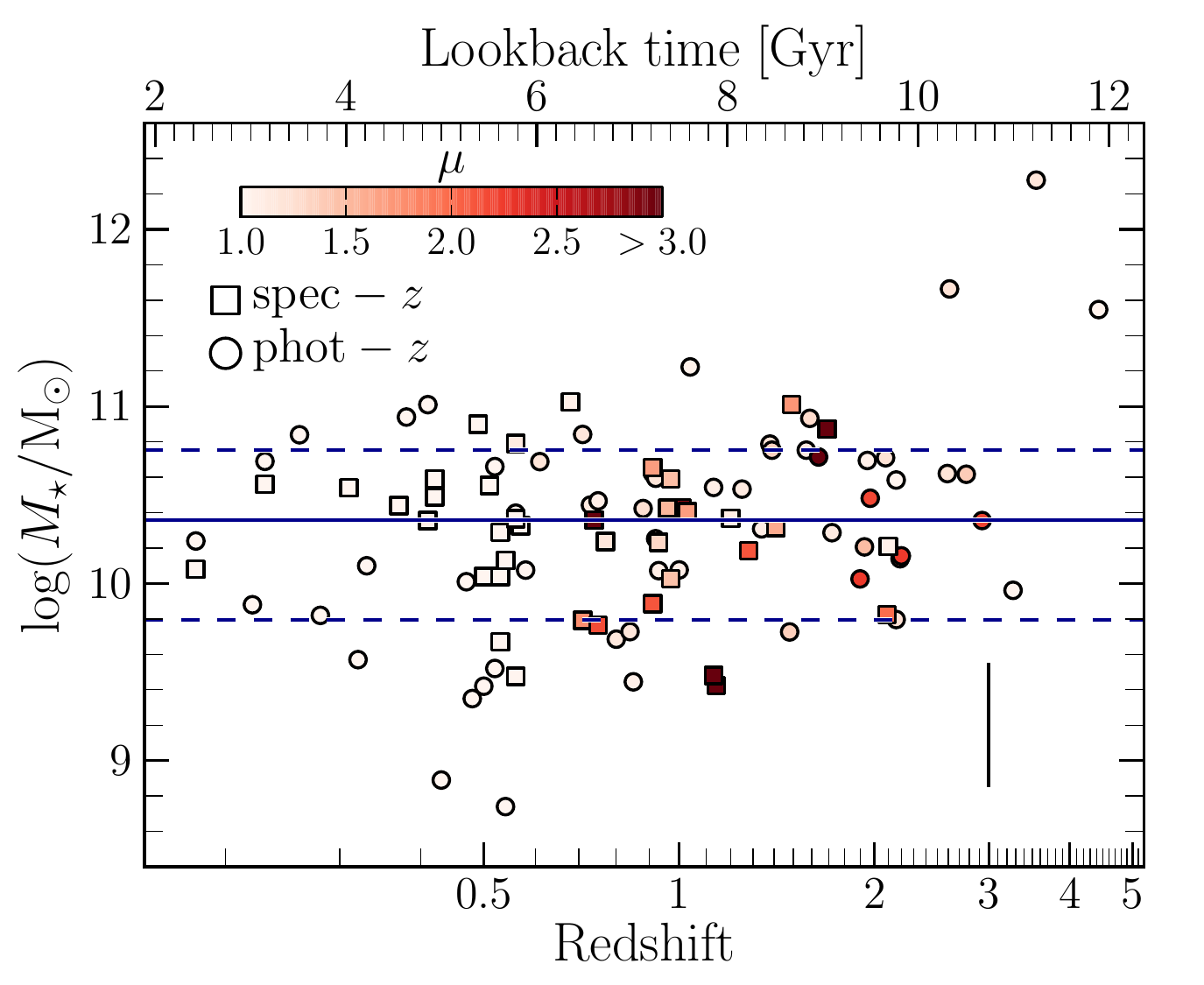}	
	\includegraphics[width=9.2cm]{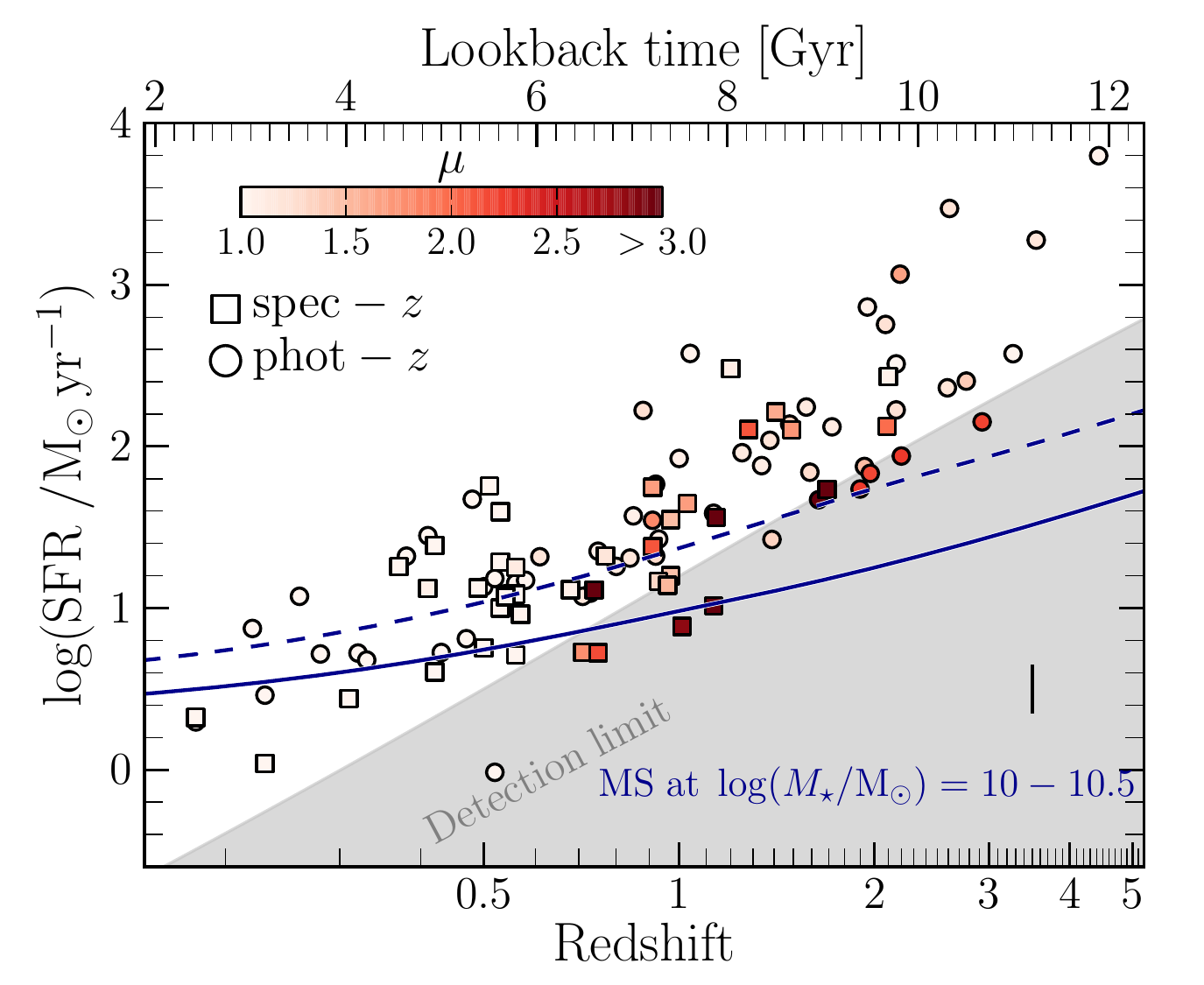}
	\caption{Lensing-corrected stellar mass ({\it left panel}) and SFR ({\it right panel}) as function of the redshift of  98 galaxies in our sample,  of which 40 have available spectroscopic redshift (squares). For the remaining 58 galaxies, we use photometric redshift estimates (circles). The data points are color-coded according to the galaxies magnification factor.  The horizontal solid and dashed lines in the left panel show the 50th and 16th/84th percentiles of the stellar mass distribution, correspondingly. Thanks to the $\mu\rm Jy$-level sensitivity of the HFF-VLA radio maps and the lensed-nature of our galaxy sample, we can detect typical, main sequence SFGs with $\log(M_\star/\rm M_\odot)=10.5$ out to $z\simeq 3$. The grey shaded region on the right panel shows our detection limit, which is given by the SFR of a galaxy detected at the $5\sigma\;(\simeq 4.5\mu \rm Jy\,beam^{-1}) $ significance level at a given redshift. The  solid and dashed line indicate the redshift evolution of the  main sequence \citep[from ][]{schreiber15} for galaxies with stellar mass of $\log(M_\star/\rm M_\odot)=10$ and $10.5$, respectively.  The moderate magnification factor ($\mu > 2$) of a minor fraction of galaxies allows us to probe lower SFRs than that SFR limit imposed by the radio maps noise.   }
	\label{fig:sample_properties}
\end{figure*}

The redshift distribution (Figure\,\ref{fig:mu_distribution}) of the final sample of 98 galaxies has a median of $z=0.9$ and 16th/84th percentiles of $z=0.4/1.9$, respectively.  In our sample, 41\% of galaxies have spectroscopic redshifts.  The median and 16th/84th percentiles of the spectroscopic redshift distribution are  $0.7^ {+0.4}_{-0.3}$.
The distribution of the stellar mass of galaxies (Figure\,\ref{fig:sample_properties}) in the sample has a median of $\log(M_\star/\rm M_\odot)=10.4^{+0.4}_{-0.6}$. The scarcely sampled mass regime below $\log(M_\star/\rm M_\odot)<10$ is a result of our radio detection limits that select massive, bright systems.   We thus expect that a mass-complete (representative) sample of radio-selected sources can be assembled  by considering galaxies with  $\log(M_\star/\rm M_\odot)\gtrsim 10$.

 \subsubsection{AGN fraction}
\label{subsub:agn_fraction}
To determine the  AGN fraction  in our  final sample, we rely on results from semi-empirical simulations \citep{wilman08} and observations of the faint radio source population \citep[e.g.,][]{bonzini13, smolcic17b}. These studies indicate that at radio flux densities below $\sim100\, \rm \mu Jy$ at $1.4-3$\,GHz,  the radio population is dominated by SFGs.  Galaxies in our sample have a median 3\,GHz flux density of just  $13^{+20}_{-5} \, \rm \mu Jy$ (Figure\,\ref{fig:mu_distribution}), where the upper/lower values correspond to the 84th/16th percentile of the flux density distribution. Based on the typical radio flux densities and predictions from \citet[][see their Figure\,14]{smolcic17b}, we expect that $\approx 75\%$ of galaxies in our sample selected at 3\,GHz are ``pure" SFGs, while the remaining fraction (i.e., 25\%) are  galaxies with some potential contribution from AGN radio luminosity. 
{   \color{black} In an attempt to identify those AGN candidates, we use  the {\it Chandra} Source Catalog Release 2.0\footnote{\url{https://cxc.cfa.harvard.edu/csc/}} \citep{evans10} and the X-ray catalog of MACSJ0717 from \citet{vanweeren16}  to find X-ray counterparts of radio sources in our sample. We find 12 positional coincidences within a 1\arcsec search radius (see Table\,\ref{table_sllsizes}). All of them have an X-ray luminosity ([$0.5-10$] keV) of $L_{\rm X}\approx10^{42.5-44.8}\,\rm erg\,s^{-1}$ that is higher than the typical limit of  $L_{\rm X} =
10^{42}\,\rm  erg\, s^{-1}$ to select X-ray AGN \citep[e.g.,][]{szokoly04}. These X-ray detected galaxies represent 11\% of the total sample, thereby, the radio emission of  galaxies in our sample is  dominated by star formation processes. We further discuss the implications of AGN-dominated sources in the relations presented in \S \ref{sec:results}.  }

\subsubsection{Radio spectral index}
\label{subsub:spectral_index}
Out of our sample of 98 field galaxies simultaneously detected in the  3\,GHz radio map and {\it HST} images, 31 of them are also detected in the 6\,GHz image. Spectral indices measured from the 3 and 6\,GHz flux densities, i.e.,  $\alpha_{\rm 6\,GHz/3\,GHz}$, are available for these sources in the VLA HFF catalogs (Heywood et al.\,{  in press}). Note that the low fraction of sources detected at 6\,GHz is mainly due to
the fainter radio flux density of SFGs at 6\,GHz, as well as the  smaller area of the 6\,GHz maps that do not fully cover the  HFF parallel fields.   
We derive a median  $\alpha_{\rm 6\,GHz/3\,GHz}$ of   $0.68^{+0.28}_{-0.46}$. 
This value is  consistent  with the typical radio spectral index of radio-selected SFGs of $\alpha\approx 0.7$ \cite[e.g.,][]{condon92, smolcic17}.

\subsection{Deriving star formation rates from radio/UV emission}\label{subsec:deriving_sfr}
To derive total SFRs from the  radio (dust-unobscured) and  UV (dust-obscured) emission of galaxies, we employ the SFR calibrations from \citet{murphy11, murphy12, murphy17} who adopt a Kroupa IMF. We normalize such SFR calibrations to the Chabrier IMF, to be consistent with the IMF employed by \citep{shipley18} to derive the stellar masses used here. Note that  the Chabrier IMF leads to SFR and stellar mass estimates that are  only $\approx 6$\% lower than those values inferred from a Kroupa IMF.   

Using the  locally measured IR-radio correlation\footnote{We note that adopting a redshift-dependent IR-radio correlation \citep[e.g.,][]{delhaize17, algera20} does not affect the sense (nor significance) of the trends reported in this work.} from \citet[][i.e., $q_{\rm IR}= 2.64$]{bell03},  the unobscured SFR of a galaxy can be estimated from its 1.4\,GHz spectral luminosity, $L_{\rm 1.4\,GHz}$, as \citep{murphy17}: 

\begin{equation}
    \label{eq:sfr_from_radio}
   \left(  \frac{\rm{SFR_{3\,GHz}}}{\rm M_\odot\,yr^{-1}} \right)= 4.87 \times 10^{-29}\; \left( \frac{L_{\rm 1.4\,GHz}}{\rm erg\,s^{-1}\,Hz^{-1}}\right).
\end{equation}
\noindent
$L_{\rm 1.4\,GHz}$ is given by
\begin{equation}
  L_{\rm 1.4\,GHz}=\frac{4\pi D_{\rm L}(z)^2}{(1+z)^{1-\alpha}} \left( \frac{1.4}{3}\right)^{-\alpha} S_{\rm 3\,GHz},
\end{equation}
\noindent
where $S_{\rm 3\,GHz}$ is the 3\,GHz flux density in $\rm erg\,s^{-1}\,cm^{-2}\,Hz^{-1}$,  $D_{\rm L}$ is the luminosity distance in cm,  and $\alpha$ is the spectral index of the power-law radio emission given by $S_{\nu} \propto \nu^{-\alpha}$. As mentioned in \S \ref{subsub:spectral_index}, an estimate of $\alpha$ is available for 31 sources with 6\,GHz counterpart. We estimate the $\rm SFR_{3\,GHz}$ of the remaining sources by adopting the typical spectral index of 3\,GHz radio-selected SFGs of $\alpha=0.7$ \citep[e.g.,][]{smolcic17}.
This simplification introduces further uncertainties to our $\rm SFR_{3\,GHz}$ estimates, which we quantify by using the 31 galaxies with available 6\,GHz counterpart to compare their $\rm SFR_{3\,GHz}$ derived from $\alpha_{\rm 6\,GHz/3\,GHz}$ and  $\rm \alpha=0.7$. We find that both estimates are well-correlated, with a dispersion of only $30\%$. We hence include an additional systematic error of 30\% to our single-band $\rm SFR_{3\,GHz}(\alpha=0.7)$ estimates.

The unobscured SFR of galaxies is estimated using the  rest-frame  far-UV (FUV) luminosity at 1600\AA, $L_{\rm FUV}$, which is sensitive to young (unobscured) stellar populations \citep[up to ages around 8\,Myr; e.g.,][]{cervino16}. $L_{\rm FUV}$ is estimated from the  FUV magnitude obtained with {\sc eazy} and reported in the HFF DeepSpace catalogs \citep{shipley18}. Then, the FUV-based SFR is estimated from the (extinction-corrected) FUV luminosity as \citep{murphy12}:
\begin{equation}
   \label{eq:fuv_sfr}
   \left(  \frac{\rm{SFR_{FUV}}}{\rm M_\odot\,yr^{-1}} \right)= 4.17 \times 10^{-44}\; \left( \frac{L_{\rm FUV}}{\rm erg\,s^{-1}}\right).
\end{equation}
\noindent

\begin{figure}
 	\begin{centering}
 		\includegraphics[width=8.5cm]{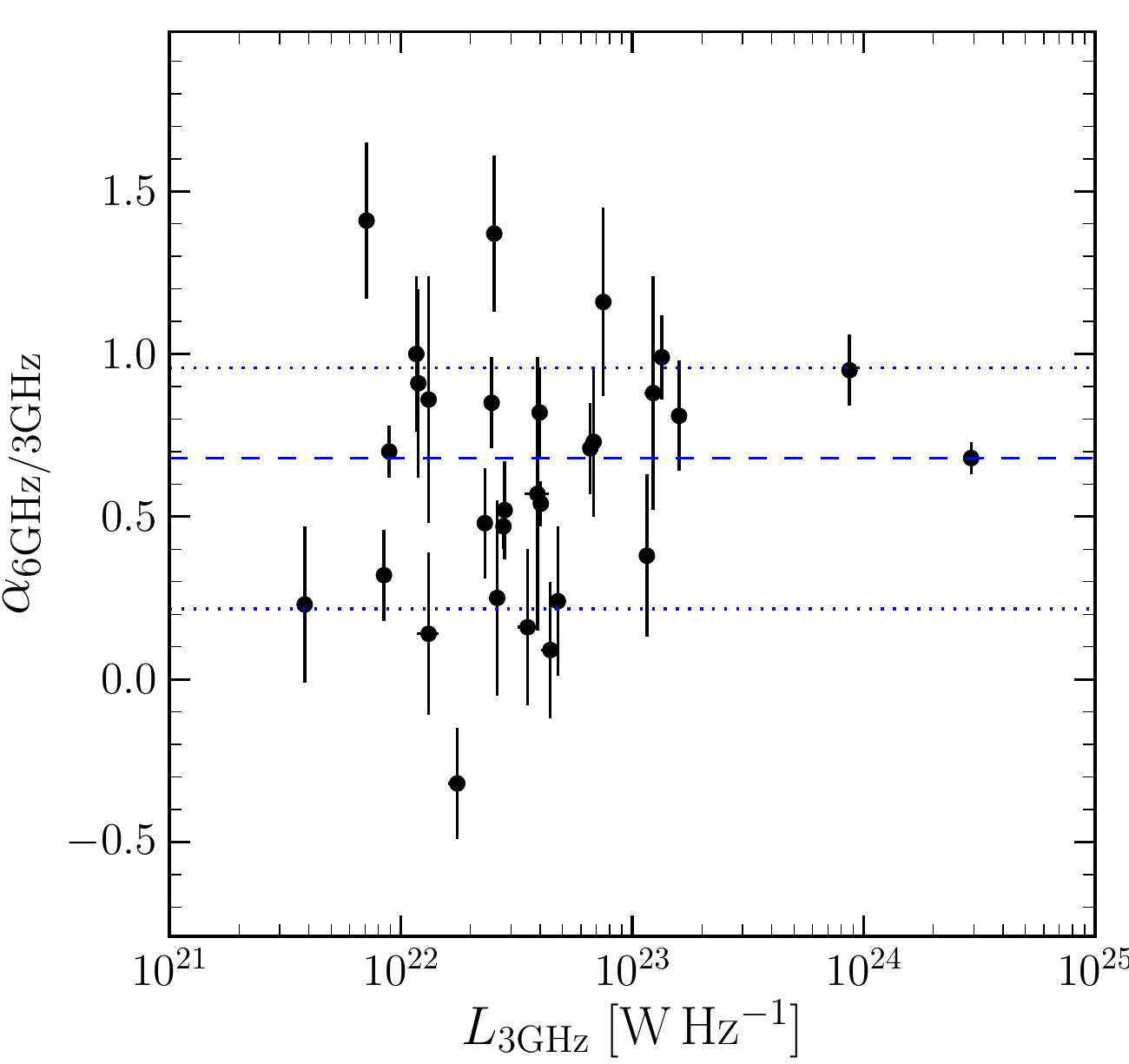}	
 		\caption{Radio spectral index as a function of 3\,GHz radio luminosity of  31 galaxies in our sample detected at both 3\, and 6\,GHz radio maps.  The horizontal dotted line shows the median spectral index of 0.68, while the dashed lines show the 16th and 84th percentiles. The median spectral index of these galaxies is consistent with the typical value of radio-selected  SFGs of 0.7. }
 		\label{fig:alpha_comparison}
 	\end{centering}
 \end{figure}

In Equation \ref{eq:fuv_sfr},  $\rm SFR_{FUV}$ is computed using the FUV as observed (i.e, it has not been corrected for internal dust extinction) to only account for the unobscured component. 
Finally, the total, lensing-corrected SFR of galaxies can be estimated as
\begin{equation}
    \rm{SFR=(SFR_{3\,GHz}+SFR_{FUV})/\mu},
\end{equation}
\noindent
where $\rm SFR_{3\,GHz}$ accounts for the population of young stars that is totally obscured by dust, and $\rm SFR_{FUV}$ reflects the contribution of unobscured  young stars. Comparing the $\rm SFR_{3\,GHz}$  with $\rm SFR_{FUV}$ (see Appendix \ref{subsec:sfr_tracers}), we find that radio-based SFRs account for $\gtrsim95\%$ of the total SFR of massive SFGs with $\log(M_\star/\rm M_\odot)>10$. Using FUV emission that is not corrected for dust attenuation can thus lead to SFRs that are underestimated, on average, by one order of magnitude. Likewise, we find that the optical-infrared (OIR) SFRs from SED fitting (reported in the HFF-DeepSpace catalogs) 
are  unreliable, showing no correlation and one order of magnitude scatter with the radio-based SFRs  (see Appendix \ref{subsec:sfr_tracers}). These results strengthen the approach of adopting $\rm SFR_{3\,GHz}$, combined with $\rm SFR_{FUV}$, to better trace the total energetic budget arising from massive star formation in galaxies.

In Figure\,\ref{fig:sample_properties}, we present the total, lensing-corrected SFRs and stellar mass as a function of the redshift of 98 field galaxies in our sample.   The  noise of $0.9\,\rm\mu Jy\,beam^{-1}$ of the 3\,GHz VLA HFF maps enables  detection of ``typical'' SFGs on the main sequence \citep[e.g.,][]{schreiber15} out to $z\approx 3$ with  $\log(M_\star/\rm M_\odot)\approx 10.5$, which is the median stellar mass of galaxies in our sample. Figure\,\ref{fig:sample_properties} also illustrates the benefits of selecting galaxies towards the HFF. By detecting intrinsically faint sources that without gravitational lensing would not have been detected with a $\rm0.9\,\mu Jy \, beam
^{-1}$ sensitivity, we are able to better sample the population of massive SFGs with $\log(M_\star/\rm M_\odot)> 10$ out to $z\approx3$.


\subsection{Estimating  the radio size of galaxies}

Heywood et al. ({  in press}) have reported the deconvolved FWHM ($\theta_{\rm M}$, and its associated uncertainty, $\sigma_{\theta_{\rm M}}$) of galaxies in the VLA Frontier Fields, including results from the 3\,GHz and 6\,GHz imaging.  Sources have been considered reliably resolved if they satisfy the condition $\phi_{\rm M}-\theta_{\rm 1/2}   \geq 2 \sigma_{\phi_{\rm M}}$; where $\phi_{\rm M}$ and $\sigma_{\phi_{\rm M}}$ are  the major axis FWHM of the source before deconvolution and its associated error (provided by {\sc PyBDSF}), and $\theta_{\rm 1/2}$ the FWHM of the beam projected along with the source PA (see Appendix A \,of Heywood et al.\,{  in press}). For sources that are not confidently resolved, here we assign  $\theta_{\rm M}+\sigma_{\theta_{\rm M}}$ as an upper limit of their deconvolved FWHM.

To compare the radio size of galaxies at 3\,GHz and 6\,GHz, we convolve the 6\,GHz images to match the beam size and shape of the 3\,GHz images (Figure\,\ref{fig:stamps}). This procedure is critical to minimize  systematic effects when comparing the radio size of compact sources  at different spatial resolutions. For example,  the total flux density and, hence, the size of moderately resolved sources can differ between radio maps with different spatial resolutions \citep{murphy17, bondi18}.  We then use {\sc PyBDSF} to perform the source extraction and size measurement  with the same parameters used by Heywood et al.\,({  in press}): a $S/N$ threshold of 5 to detect peaks of emission and a secondary $S/N$ threshold of 3 to identify the pixels  (islands of emission) associated  to those peaks. The errors and upper limits to the radio size  from these convolved images at 6\,GHz are derived as in Heywood et al.\,({  in press}). This new catalog is combined with information from the 3\,GHz imaging from Heywood et al.\,({  in press}) to get a matched-resolution ($\approx0\farcs7$) catalog containing the radio size (and error) estimates of galaxies in the sample at 3\,GHz and 6\,GHz. Around 50\% of the 3 and 6\,GHz radio sources in our sample are reliably resolved: 53 out  of the 98 at 3\,GHz, and 12 out of 31 at 6\,GHz.  

Finally, since our sources are only marginally resolved (i.e., $R_{\rm eff-rad} \lesssim \theta_{\rm 1/2}$, see an example in Figure\,\ref{fig:stamps}), the effective radii of the radio sizes is well approximated by $R_{\rm eff-rad} \approx \theta_{\rm M}/2.430$ \citep{murphy17}, where $\theta_{\rm M}$ is the deconvolved FWHM provided by {\sc PyBDSF}. In applying this approximation we assume that most galaxies in the sample are exponential disks, which is consistent with the S\'{e}rsic index of $n\approx 1$ reported for typical SFGs  \citep[using stellar emission; ][]{nelson16} and even highly-active SFGs at $z\gtrsim2$  \citep[using dust continuum emission;][]{hodge16, hodge19, gullberg19, tadaki20}. This approximation might deviate from the true effective radius of galaxies that have cuspier light profiles, including a fraction of compact quiescent and starburst galaxies \citep[e.g., ][]{wuyts11}. However, even with an angular resolution that is three times better than ours, \citet{elbaz18} finds no significant changes in the effective radius of $z\sim2$ compact galaxies (from dust continuum emission) if  S\'ersic indices higher than one are used. 
Finally, the effective radius is corrected for lensing magnification as follows: $R_{\rm eff-rad}^{\rm cor}=R_{\rm eff-rad}/\sqrt{\mu}$. For simplicity, in the following, we refer to this lensing-corrected effective radius as $R_{\rm eff}$.


\subsection{Estimating  the UV/optical  size of galaxies}

We use the  Python package  {\tt statmorph}\footnote{\url{https://statmorph.readthedocs.io/en/latest/}} \citep{rodriguezgomez18} to derive the UV/optical effective radius  of galaxies in the sample.  This radius is estimated for each of the six {\it HST} bands used in this work (Figure\,\ref{fig:stamps}), but only for those galaxies flagged with good photometric data (flag=0) in the HFF DeepSpace catalog \citep{shipley18}. Most galaxies ($>95\%$) profit from robust photometric imaging in the {\it HST}/ACS bands, whereas around $60\%$ of them are also covered by the {\it HST}/WFC3 imaging. 

 To implement {\tt statmorph}, we first create $15\arcsec\times15\arcsec$\, width cutouts per each of the six {\it HST} photometric bands. Then, the background  is estimated (and  subtracted from the original image) through sigma-clipped statistics  with the {\tt photutils} Python library.  We independently create segmentation maps for the Brightest Cluster Galaxies (BCGs)  as these are not available in the HFF DeepSpace project. This is done with the {\tt photutils.SegmentationImage} and {\tt photutils.deblend-sources} libraries that enable us to deblend the emission of the BCGs and neighboring sources. 

The effective (half-light)  radius is estimated by using elliptical apertures centered  at the asymmetry center of the galaxy emission. The outer radius containing the total flux of the galaxy, $R_{\rm max}$, is defined  in {\tt statmorph} as the distance between the asymmetric center of the galaxy emission and the most distant pixel within the detection mask.  This mask is derived by smoothing the galaxy image with a  $3\times3$ boxcar kernel and considering the contiguous group of pixels that are $1\sigma$ above  the mode \citep{rodriguezgomez18}. We neglect  the size estimates  that are flagged by {\tt statmorph} as not physically meaningful quantities. This type of results represents less than 10\% of total outputs.  
Finally, to consider the effect of the  PSF on our $R_{\rm eff}$ estimates, we apply  the relation $R_{\rm eff}^{\rm intrinsic}=(R_{\rm eff}^2-R_{\rm PSF}^2)^{1/2}$; where  $R_{\rm PSF}$ has been derived following the same non-parametric analysis on the PSF images, and $R_{\rm eff}^{\rm intrinsic}$ is the PSF-corrected effective radius.   The lensing magnification is also corrected by evaluating: $R_{\rm eff}^{\rm cor}=R_{\rm eff}^{\rm intrinsic}/\sqrt{\mu}$. In the following, we simply refer to the PSF- and lensing-corrected effective radius as $R_{\rm eff}$.

The code {\tt statmorph} also fits 2D Sérsic models to the galaxy emission from which the effective radius, $R_{\rm eff}^{\rm Sersic}$, is also derived. We verified that $R_{\rm eff}^{\rm Sersic}$ and the non-parametric $R_{\rm eff}$ estimates  are consistent.  The $R_{\rm eff}-R_{\rm eff}^{\rm Sersic}$ relation shows a dispersion of $\sim 0.2$ dex.  In the case of extended galaxies, however,  $R^{\rm Sersic}_{\rm eff}$  tends to be higher than $R_{\rm eff}$  as a 2D Sérsic model fails to reproduce the clumpy structure of extended, well-resolved galaxies in our sample.

\subsubsection{Wavelength dependence of the galaxy size in the  UV-to-optical  regime}\label{subsec:wave-vs-hstsize}

\begin{figure*}
	\begin{centering}
		\includegraphics[width=18cm]{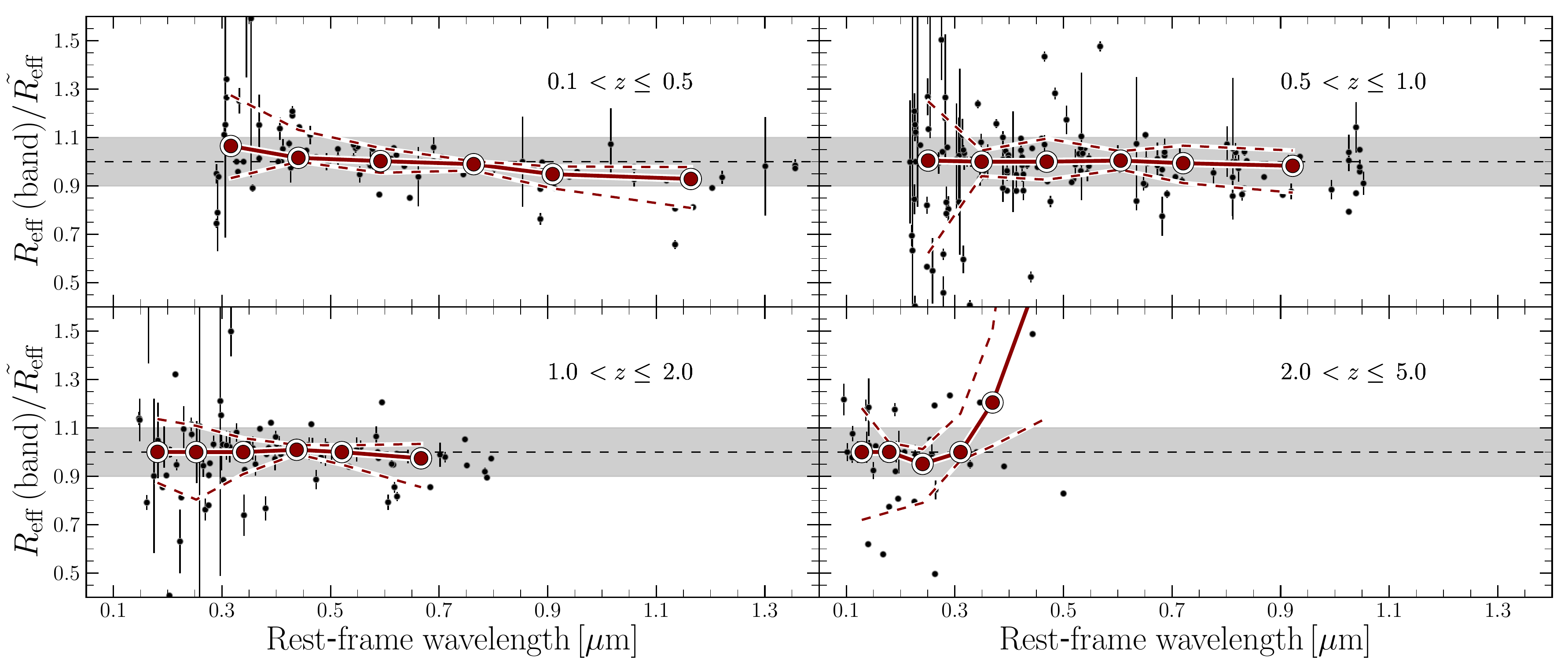}
		\caption{The effective radius of galaxies at a fixed rest-frame wavelength, $R_{\rm eff} ({\rm band})$, measured from the six {\it HST} bands. The values in the ordinate have been normalized to the median effective radius ($\tilde{R}_{\rm eff}$) from the six {\it HST} bands. Four different redshift bins are presented.  The grey circles show the $R_{\rm eff} ({\rm band})/\tilde{R}_{\rm eff}$ ratio for all the galaxies in the sample and for all the {\it HST} bands used here. The magenta circles show the median $R_{\rm eff} ({\rm band})/\tilde{R}_{\rm eff}$ ratio per {\it HST} band, while the dashed blue lines show their associated dispersion (16th and 84th percentiles).  The grey shaded region illustrates the  $\pm 10\%$ variance with respect to the median effective radius from   the six {\it HST} bands. The dashed horizontal line is set at $R_{\rm eff} ({\rm band})/\tilde{R}_{\rm eff}=1$. This plot demonstrates that the effective radius of most galaxies in the sample remains constant across the  UV-to-optical regime.   }
		\label{fig:hstsize_wave}
	\end{centering}
\end{figure*}

To investigate the UV/optical size evolution of galaxies across cosmic time, we need to consider the  effects of the morphological $k$-correction \citep[e.g., ][]{conselice14}.   As the redshift increases,   the ACS and WFC3 filters probe different regimes of the galaxy spectrum and hence distinct stellar populations. Without correcting for this effect, the inferred size evolution of galaxies might simply reflect the  dependence of the galaxy size on the wavelength. 
To explore such a correlation, we present in Figure\,\ref{fig:hstsize_wave} the effective radius measured from the ACS and WFC3 imaging as a function of the rest-frame wavelength. For each galaxy, the effective radius derived from the different {\it HST} filters, $R_{\rm eff} (\rm band)$,  has been normalized to the median value $\tilde{R}_{\rm eff}$.
In all but the highest redshift bin, we find  that the size of galaxies varies, on average, less than 10\% across all the  {\it HST} filters used in this study.  This is consistent with the smooth evolution of the UV-to-optical size of galaxies previously reported in the literature \citep[e.g., ][]{vanderwel14,  vika15, ribeiro16}. At $z>2$, we observe a steep increment of the galaxy size at longer wavelengths; however, this trend is driven by a small number of data points and hence it is associated with a high degree of uncertainty. Finally, we verify that the wavelength dependence of the galaxy size in the  UV-to-optical  regime (Figure\,\ref{fig:hstsize_wave})  does not significantly change if galaxies with different mass and SFR are considered.

Given the smooth variations of galaxy size throughout UV-to-optical wavelengths,    a single measurement over  $0.1-0.4\,\mu$m and $0.4-1.0\,\mu$m can  provide an estimate for the size of galaxies in the rest-frame UV and optical regime, respectively. Therefore, we  group all the available size measurements falling within the UV ($0.1-0.4\,\mu$m) and optical ($0.4-1.0\,\mu$m) rest-frame wavelength ranges, and  report the corresponding median value as the UV and optical size of galaxies.

\subsection{Strengths and limitations of our  size measurements}

Comprehensive simulations that investigate the biases and uncertainties in measuring the  size of galaxies from UV/optical imaging have been performed \citep[][]{meert13,  mosleh13,  davari14}. The consensus is that either parametric (e.g., fitting S\'ersic profiles) or  non-parametric methods can retrieve the size of galaxies  with $S/N\gtrsim10$   without introducing strong systematic errors (up to 20\%). Past studies also suggest that the cosmological surface brightness dimming does not significantly affect the UV/optical size measurement of galaxies out to $z\sim4$ \citep{paulino-afonso16, ribeiro16}. Particularly, using simulations of {\it HST} observations,  \citet{davari14} find that there are no significant biases in the effective radius inferred from the cumulative radial flux distribution of galaxies --  as those derived here. 
 For galaxies with $R_{\rm eff}\gtrsim 0\farcs1$, as in our sample, the UV/optical sizes can be measured without introducing strong biases and with $\lesssim 5\%$ uncertainties \citep[][see their Figure\,3]{davari14}.

However, because high-redshift galaxies exhibit faint radio emission that is centrally concentrated 
\citep[e.g., ][]{murphy17, bondi18, jimenezandrade19},  measuring the radio size is challenging even with the deep, high-resolution (sub-arcsec) radio observations delivered by the HFF VLA survey. Contrary to the UV/optical emission that is spatially resolved across multiple resolution elements, our radio sources are slightly resolved, which could lead to systematic uncertainties in our size measurements. To verify this, we infer  $R_{\rm eff}$ from the $S_{\rm peak}/S_{\rm int}$ ratio for a circular exponential disk observed with a circular Gaussian beam. This ratio is given by \citep[see Appendix C of][]{murphy17}:
\begin{equation}
     \frac{S_{\rm peak}}{S_{\rm int}}=2z^2 \left[1- \sqrt{\pi}  z \exp(z^2) {\rm erfc}(z)\right],
     \label{eq:speak-sint}
 \end{equation}
\noindent
where $z\approx 0.50398 \left( \theta_{1/2}/R_{\rm eff} \right)$, with $\theta_{1/2}$ the FWHM of a circular beam\footnote{Due to the elliptical beam of our radio images, we consider  $\theta_{1/2}$ as the geometrical mean of the FWHM of the major and minor axis of the synthesized beam.}.  While Equation \ref{eq:speak-sint} is valid for a circular (face-on) exponential disk, the $S_{\rm peak}/S_{\rm int}$ ratio of an elliptical (inclined) exponential disk can be estimated as
\begin{equation}
    \frac{S_{\rm peak}}{S_{\rm int}} \approx \left[  \left(\frac{S_{\rm peak}}{S_{\rm int}}\right)_{\rm M}  \left(\frac{S_{\rm peak}}{S_{\rm int}}\right)_{\rm m}  \right]^{1/2}, 
         \label{eq:speak-sint_2}
\end{equation}
\noindent
 where $(S_{\rm peak}/S_{\rm int})_{\rm M}$ and $(S_{\rm peak}/S_{\rm int})_{\rm m}$ are the respective $S_{\rm peak}/S_{\rm int}$ ratios along the major (M) and minor (m) axes of an elliptical disk.

\begin{figure*}
 	\begin{centering}
 		\includegraphics[width=18.15cm]{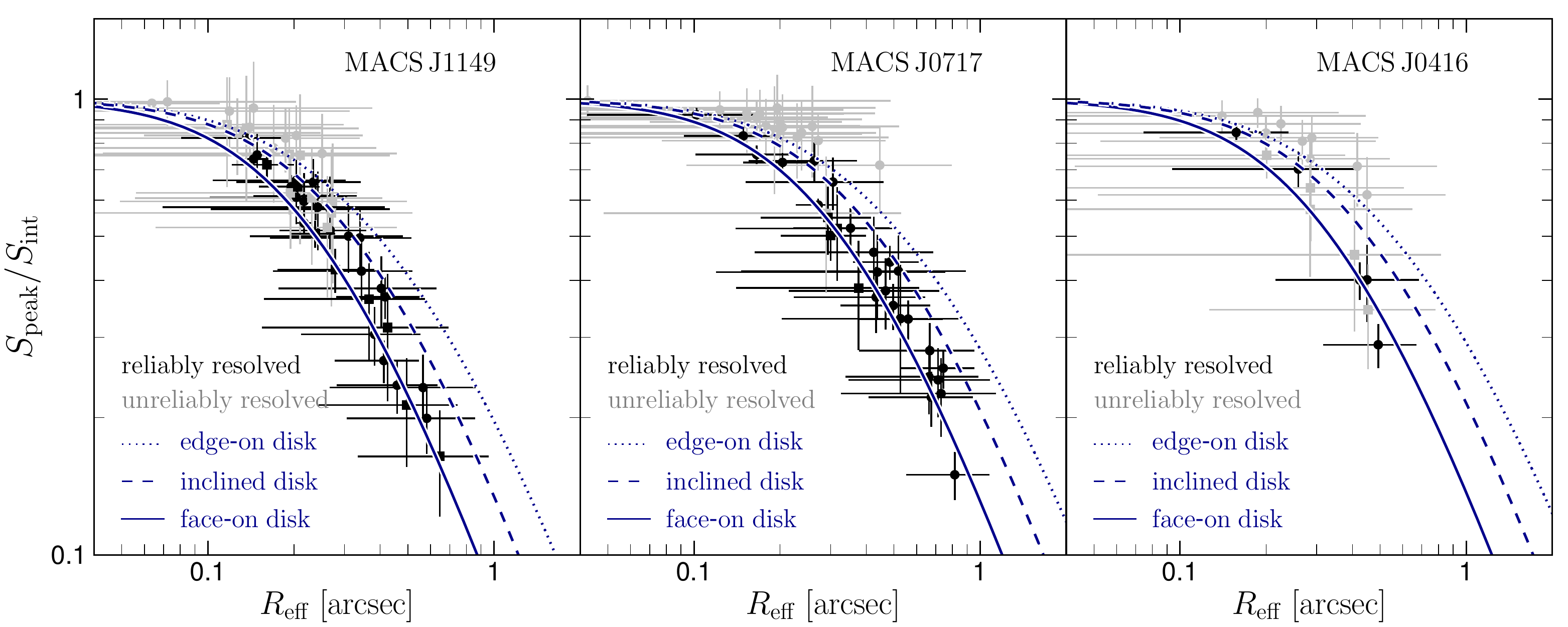}	
 		\caption{  
 		 The ratio of peak-to-integrated flux densities, $S_{\rm peak}/S_{\rm int}$, of sources in our sample as a function of the deconvolved disk effective radius, $R_{\rm eff}$, in units of arcsec. We present the $S_{\rm peak}/S_{\rm int}-R_{\rm eff}$ relation for SFGs in the MACS\,J1149, MACS\,J0717, and MACS\,J0416 fields separately, to account for the different spatial resolution of their respective VLA radio image.  The black circles (squares) show the reliably resolved sources at 3\,GHz (6\,GHz), while the grey symbols represent the unreliably resolved sources. The expected $S_{\rm peak}/S_{\rm int}-R_{\rm eff}$ relation for a a circular (face-on) exponential disk is shown by the solid blue line (Equation \ref{eq:speak-sint}). The dashed blue lines shows the  $S_{\rm peak}/S_{\rm int}-R_{\rm eff}$ relation for an inclined disk [$\cos(i) = 0.5$, with $i$ the inclination angle], whereas the dotted blue line corresponds to a nearly face-on disk [$\cos(i) = 0.25$]. This demonstrates that the  size of reliably resolved sources meets expectations from  their $S_{\rm peak}/S_{\rm int}$ ratio.    }
 		\label{fig:test_size_1}
 	\end{centering}
 \end{figure*}

In Figure\,\ref{fig:test_size_1}, we present the observed and expected  $S_{\rm peak}/S_{\rm int}-R_{\rm eff}$ relation of sources in the MACS\,J1149, MACS\,J0717, and MACS\,J0416 fields. We find that the effective radius of reliably resolved sources lies between the expected value for an edge-on and face-on disk. Such an agreement, between the observed and expected  $S_{\rm peak}/S_{\rm int}-R_{\rm eff}$ relation, strengthens the reliability of our size measurements via 2D Gaussian fitting with {\sc PyBDSF}. 

\subsection{  Maximum recovered radio size} 
\label{subsub:selection_function}
\begin{figure}
 	\begin{centering}
 		\includegraphics[width=8.6cm]{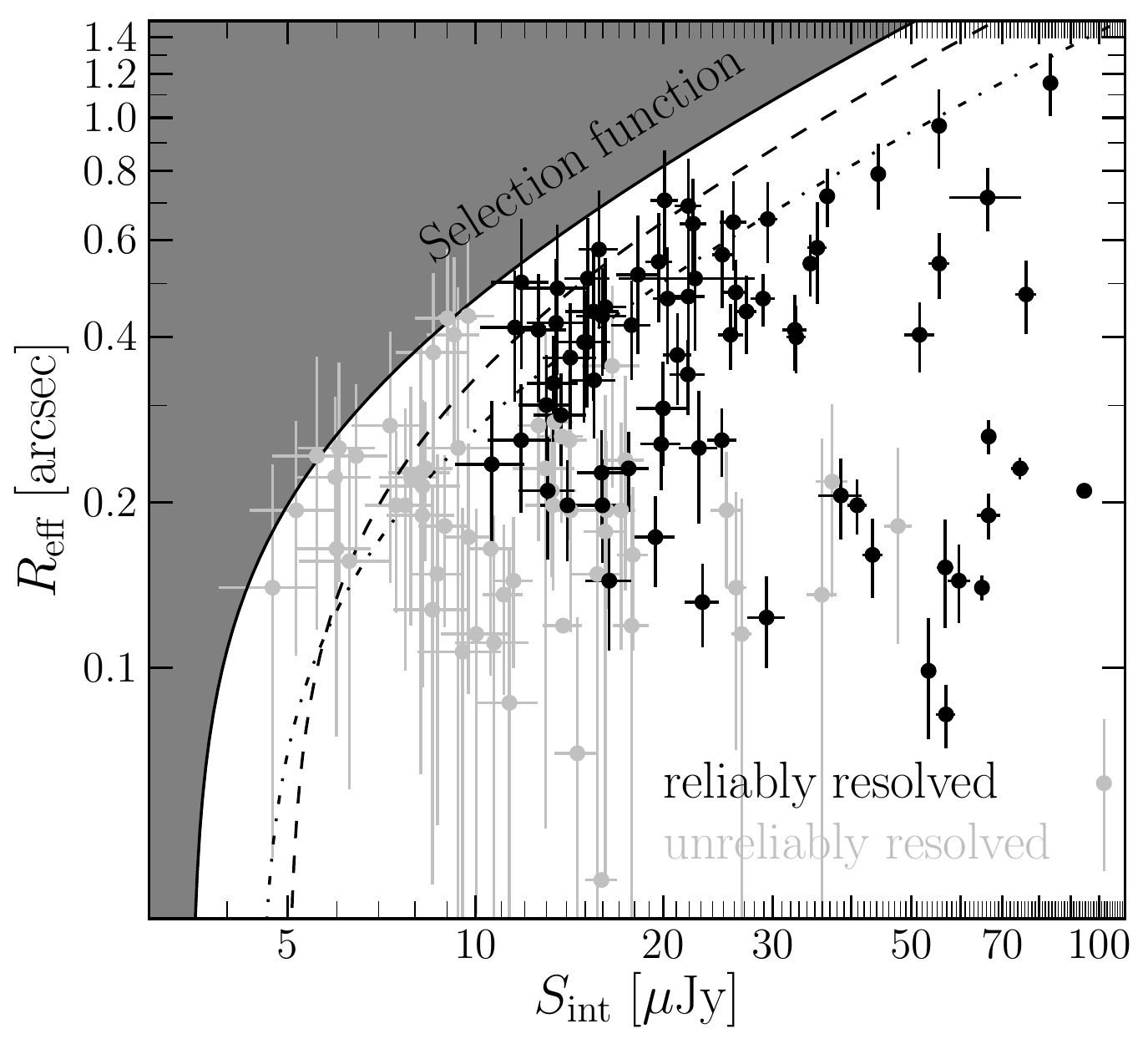}	
 		\caption{  Maximum detectable angular size as a function of total flux density of sources in the HFF radio images: MACSJ0416 (dashed line), MACSJ0717 (solid line), MACSJ1149 (dash-dotted line).  Reliably resolved sources are shown as black circles, while those unreliably resolved are presented in grey. Close to the detection limit, our selection function bias the sample toward compact sources that can not be reliably resolved.   }
 		\label{fig:selection_function}
 	\end{centering}
 \end{figure}

{  
We use Equations \ref{eq:speak-sint} and \ref{eq:speak-sint_2} to infer the maximum detectable angular size as a function of flux density of sources in our sample. This allows us to derive the maximum radio size of galaxies that can be detected at a given redshift, stellar mass, and SFR.  To this end, we solve for $R_{\rm eff}$ in Equation \ref{eq:speak-sint} using the Newton-Raphson method with the {\tt scipy.optimize} library in Python. We then evaluate $R_{\rm eff}(S_{\rm int})$ using the respective FWHM of the major and minor axis of the beam and rms noise level from the three different HFF radio images (see Figure\,\ref{fig:selection_function}). The selection function imposed by the resolution and sensitivity of the MACSJ0416, MACSJ0717, and MACSJ1149 radio images is consistent: galaxies with flux densities below 10$\mu Jy$, close to the detection limit, are preferentially detected when they are compact. Also, due to their faint nature, such compact sources can not be reliably resolved.  We  discuss the effects of this selection function in the results presented in  \S\ref{subsec:size-mass} and \ref{subsec:size-starformation}. }

\section{Results and Discussion}
\label{sec:results}

 In this Section, we  compare the radio size of galaxies measured at 3 and 6\,GHz (\S \ref{subsec:hst-vs-vlasize}) and investigate the  dependence of the galaxy size on  stellar mass (\S \ref{subsec:size-mass}),  star formation activity (\S \ref{subsec:size-starformation}), and redshift (\S \ref{subsec:size-redshift}). The redshift, SFR, stellar mass, and multi-wavelength size estimates of all the galaxies in our sample are presented in Table \ref{table_sllsizes}.

\subsection{Comparing the radio size of galaxies at 3\,GHz and 6\,GHz}\label{subsec:hst-vs-vlasize}

\begin{figure}
	\begin{centering}
		\includegraphics[width=8.7cm]{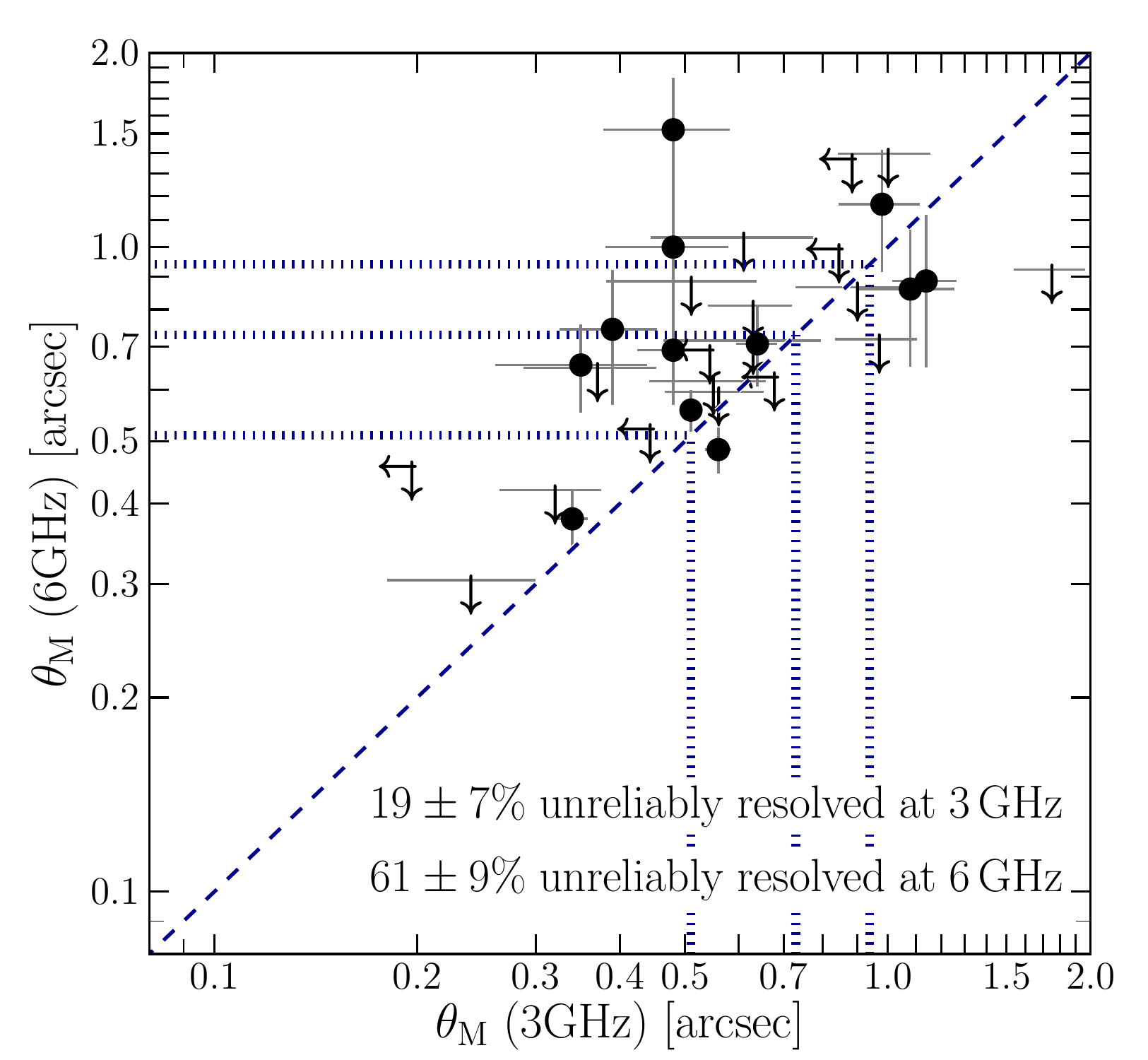}	
		\caption{Deconvolved  FWHM ($\theta_{\rm M}$) of the radio continuum emission of galaxies in 31 galaxies  at 6\,GHz as a function of the one measured at 3\,GHz. The deconvolved FWHM at 6\,GHz has been measured from the map that has been convolved to map the resolution of the 3\,GHz map. The black circles show the galaxies for which both their sizes at 3 and 6\,GHz are reliable resolved. The horizontal/vertical arrows are upper limits to the deconvolved FWHM at 3\,GHz/6\,GHz of sources that are not reliably resolved. The dashed diagonal line represents the  one-to-one relation. The dotted vertical/horizontal lines correspond to the synthesized beam of the three VLA Frontier Fields images from which the 3 and 6\,GHz size measurements are obtained.  Even with a $\sim0\farcs7$ resolution, 19\% (61\%) of sources in our sample are not reliably resolved at 3\,GHz (6\,GHz). The larger fraction of sources unreliably resolved at 6\,GHz hints toward the more compact radio size of galaxies at higher frequencies.   }
		\label{fig:radiosize-compa}
	\end{centering}
\end{figure}

In Figure\,\ref{fig:radiosize-compa}, we compare the radio sizes of galaxies at 3\,GHz and 6\,GHz from our VLA Frontier Fields images. Out of the initial sample of 98 radio-selected galaxies, 31 of them are detected in the matched-resolution radio images at 3\,GHz and 6\,GHz. Among these common sources, 25 are reliably resolved  at 3\,GHz, and only 12 at 6\,GHz.  As a result,  our data set is dominated by upper limits to the size of galaxies that are not reliably resolved at 6\,GHz. To derive  median properties  of  this censored data set, we use survival analysis via the  \citet[KM; ][]{kaplan58} estimator with the Python package {\tt lifelines}\footnote{\url{https://lifelines.readthedocs.io/en/latest/Survival\%20Analysis\%20intro.html}}. This statistical technique employs the censored sample, where upper limits are present, to  provide a maximum-likelihood-type reconstruction of the true distribution function \citep{feigelson85}. 

We  derive a median 3\,GHz size of $R_{\rm eff}=1.3\pm0.3 \rm \, kpc$ for all the 98 galaxies in our sample,  of which 53 are reliably resolved. For the 31 sources with available 3 and 6\,GHz counterparts, we  find a median   $R_{\rm eff}$ of $1.3\pm0.3\rm \, kpc$ at 3\,GHz and  $1.1^{+0.7}_{-0.3} \rm \, kpc$ at 6\,GHz\footnote{Out of the sample of 98 galaxies selected at 3\,GHz, 23 of them are detected above the 5$\sigma$ level in the original 6\,GHz images at native resolution ($\approx 0\farcs35$). Only 6 of them are reliably resolved. We derive an upper limit to their median size of $R_{\rm eff}<1.4$\,kpc. The median 6\,GHz size of 31 SFGs detected in the convolved maps of $1.1^{+0.7}_{-0.3} \rm \, kpc$ is, therefore, consistent with the upper limit inferred from 23 SFGs detected in the native resolution images.}, where the upper/lower limits correspond to the 95\% confidence interval for the median.  Using Monte Carlo error propagation, we find that that the 3\,GHz size is a factor $1.1\pm0.1$ larger than the 6\,GHz size of galaxies in the sample.

The radio continuum size of galaxies at 3\,GHz has been previously explored. In the COSMOS field, \citet{bondi18} \& \citet{jimenezandrade18} report a median effective radius of $1.5\pm0.2$\,kpc  for massive SFGs $[\log(M_\star/M_\odot) \lesssim 10.5]$ over $0.3 \lesssim z \lesssim 3$; while \citet{miettinen17b} finds a median effective radius  of $1.9\pm0.2$\,kpc for 152 sub-millimeter selected galaxies  over $1<z<6$ with a median redshift of $z = 2.23\pm0.13$. Similarly, \citet{cotton18} report a median 3\,GHz effective radius of $1.0\pm0.3\,$kpc for $z\approx1$ SFGs in the Lockman-Hole. These  values are in agreement, within the uncertainties,  with the median effective radius of galaxies in our sample of $R_{\rm eff}=1.3\pm0.3 \rm \, kpc$. We attribute the scatter of these median $R_{\rm eff}$ values to the distinct selection criteria and different properties of the radio images. First, our sample might be ``contaminated'' by AGN (see \S \ref{subsec:sample}) that could lead to more compact radio sizes \citep{bondi18}. This effect could account for the $\approx 15\%$ lower effective radius of galaxies in our sample with those from ``pure'' SFGs reported by \citet{bondi18} \& \citet{jimenezandrade18}.  
Second, the maximum detectable angular size of radio sources depends on both the resolution and  sensitivity of radio maps \citep[e.g., ][]{bondi08}. Deeper, high-resolution radio surveys are more sensitive to compact radio sources.  We thus expect that the median radio size of galaxies  reported in the literature is influenced by the different depth and resolution of the radio maps.

There also exists information on the radio continuum size of galaxies at lower/higher frequencies.  While our radio sizes at 3 and 6\,GHz are two times larger than those at 10\,GHz from $z \approx 1.2$ SFGs  \citep{murphy17},  our  3 and 6\,GHz  sizes are two-to-three times smaller than those at 1.4\,GHz from $0.2\lesssim z\lesssim 2$ SFGs \citep{muxlow05, muxlow20, lindroos18}. Such trends remain even if we compare the radio size of SFGs in our sample spanning over a similar redshift (and stellar mass) range than those studied by \citet{murphy17, muxlow05, muxlow20,lindroos18}. These findings meet expectations from the frequency-dependent cosmic ray diffusion \citep[but see][who report large radio size of galaxies at 5.5\,GHz]{guidetti17}. Lower-energy electrons emitting at lower frequencies can diffuse further into the ISM of galaxies as these have longer radiative lifetimes \citep{murphy09,thomson19}. As a result, the radio size of galaxies tends to decrease with increasing frequency. We further discuss the implications of frequency-dependent cosmic ray diffusion into the observed redshift-radio size relation in \S \ref{subsec:size-redshift}.

\begin{figure*}
	\begin{centering}
		\includegraphics[width=18cm]{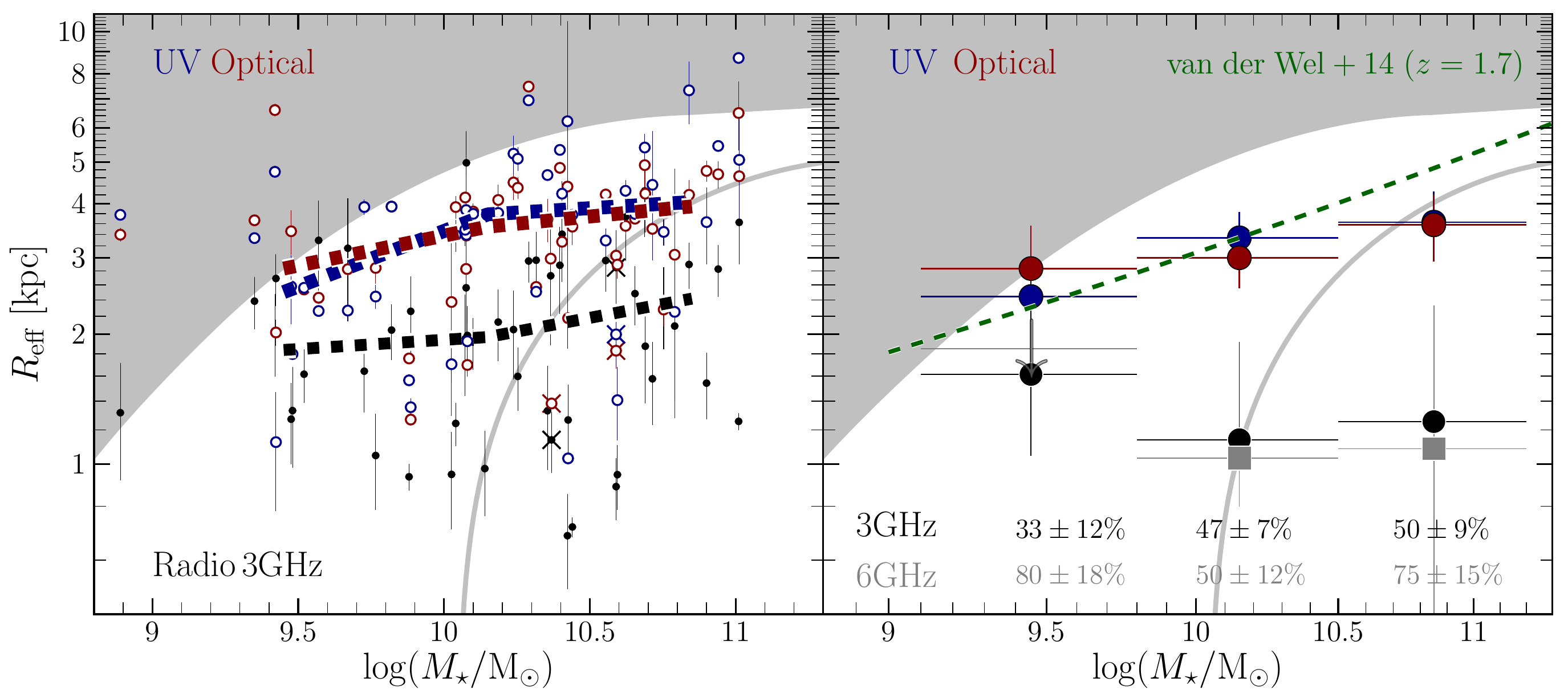}	
		\caption{
		{\it Left panel--} The radio/UV/optical effective radius of galaxies that are reliably resolved at 3\,GHz as a function of their stellar mass. The dashed lines show the median effective radius per stellar mass bin (0.75\,dex width). {   \color{black} The cross symbols denote the radio sources with a reported X-ray counterpart.} {\it Right panel--} The median radio/UV/optical effective radius of all galaxies in the sample (i.e., sources that are reliably and unreliably resolved at 3\,GHz) per stellar mass bin.   To account for the upper limits to the radio sizes of unreliably resolved galaxies,  we derive the median effective radius per stellar mass bin (and its associated uncertainty) via the KM  estimator \citep{kaplan58}. The down arrows indicate the upper limit to the median size of galaxies per  bin. The length of the error bars indicates the error of the median and the stellar mass bin, respectively. {   The grey curve shows the maximum $R_{\rm eff}$ that can be observed for a typical (main sequence) galaxy  at $z=0.9$, which is the median redshift of our galaxy sample. The grey region illustrates the parameter space that can not be probed with our radio data set.}
		The values at the bottom indicate the fraction of unreliably resolved galaxies at 3 and 6\,GHz, respectively.  In both panels,  the UV/optical size of galaxies tend to increase with stellar mass, resembling --to some extent-- the shallow slope of the stellar-mass relation of SFGs \citep[e.g.,][]{vanderwel14}. There is no evident dependence between the radio size  and stellar mass of  SFGs in our sample.  }
		\label{fig:size-mass}
	\end{centering}
\end{figure*}

\subsection{Size {\it vs} stellar mass}\label{subsec:size-mass}
 
To address the dependence of stellar mass on galaxy size, we distribute our galaxy sample into three mass bins: $9.1<\log( M_\star/\rm M_\odot)\leq 9.8$, $9.8< \log(\rm M_\star/M_\odot)\leq 10.5$, and  $10.5<\log( M_\star/\rm M_\odot)\leq11.2$. We present the size -- stellar mass relation for galaxies that are reliably resolved at 3\,GHz (left panel of Figure\,\ref{fig:size-mass}) and all  galaxies in the sample (i.e., reliably and unreliably resolved; right panel of Figure\,\ref{fig:size-mass}).

Using reliably resolved sources in the analysis, we find that the median radio size of SFGs mildly increases with stellar mass. Yet, this trend vanishes after considering unreliably resolved sources via survival analysis, indicating that there is no clear dependence between the radio size and  stellar mass of SFGs (see Table \ref{table_mediansizes}). 
 No robust constraints on the size -- stellar mass relation can be inferred from the 6\,GHz data, for which  we can only provide upper limits to the median size of galaxies at different mass bins. 
 
 Contrary to the radio, the  UV/optical extent  of  galaxies in our sample positively correlates with their stellar mass  -- as  previously reported in the literature \citep[see Figure\,\ref{fig:size-mass} and Table \ref{table_mediansizes}; e.g.,][]{morishita14, vanderwel14}. The UV/optical emission of SFGs with $10.5<\log( M_\star/\rm M_\odot)\leq11.2$ tend to be a factor $1.6\pm 0.3$ more extended than less massive SFGs with $9.1<\log( M_\star/\rm M_\odot)\leq 9.8$.  We verify that none of the trends described above change if only SFGs at high ($z\gtrsim1$) and low redshifts ($z\lesssim1$) are included in the analysis.
{   \color{black} Also, we corroborate that the  size -- stellar mass relations presented in Figure\,\ref{fig:size-mass} are not significantly affected if we exclude the 12 AGN candidates in our sample (see Figure\,\ref{fig:size-mass_noagn} in the Appendix).}
 
\begin{table*}
    \centering
		\caption{The median  radio and UV/optical size of SFGs as a function of their stellar mass, SFR, and redshift.}
		{\small
			\begin{tabular}{ c | c c c c c c c}
				\hline \hline 
				     $M_\star$, $\rm SFR$, and $z$ bin&        \multicolumn{3}{c|}{ {SFGs reliably resolved at 3\,GHz}}  &  \multicolumn{4}{|c}{{SFGs reliably and unreliably resolved at 3\,GHz}} 
				\\[0.5ex]
				  &  $R_{\rm eff}^{\rm UV}$/kpc & $R_{\rm eff}^{\rm opt}$/kpc & $R_{\rm eff}^{\rm 3\,GHz}$/kpc  & $R_{\rm eff}^{\rm UV}$/kpc & $R_{\rm eff}^{\rm opt}$/kpc & $R_{\rm eff}^{\rm 3\,GHz}$/kpc & $R_{\rm eff}^{\rm 6\,GHz}$/kpc\\[0.5ex]
				  
				  \hline 
				  
				  $9.1< \log(M_\star/ \rm M_\odot) \leq9.8$ & $2.5\pm0.8$ & $2.8\pm0.9$ & $1.8\pm0.6$ & $2.4\pm0.6$ & $2.8\pm0.7$ & $1.6^{+1.1}_{-0.6}$ & $<1.9$ \\
				  
				  $9.8< \log(M_\star/ \rm M_\odot) \leq10.5$ & $3.8\pm0.8$ & $3.5\pm0.7$ & $2.0\pm0.4$ & $3.3\pm0.5$ & $3.0\pm0.4$ & $1.1^{+0.8}_{-0.2}$& $1.0^{+1.2}_{-0.2}$  \\
				  
				  $10.5< \log(M_\star/ \rm M_\odot) \leq11.2$ & $4.0\pm1.0$ & $3.9\pm1.0$ & $2.4\pm0.6$ & $3.6\pm0.6$ & $3.6\pm0.6$ &$1.2^{+0.8}_{-1.2}$ & $1.0^{+0.6}_{-1.0}$ \\[0.5ex]
				  
				\hline 
				  $0.8< \log(\rm SFR/ M_\odot\,yr^{-1}) \leq1.6$\tablenotemark{a} & $4.3\pm0.9$ & $3.9\pm0.8$ & $2.0\pm0.5$ & $3.7\pm0.6$ & $3.6\pm0.6$ & $1.3^{+0.7}_{-0.1}$ & $1.1^{+0.7}_{-0.3}$  \\
				  
				 $1.6< \log(\rm SFR/ M_\odot\,yr^{-1}) \leq2.4$\tablenotemark{a} & $3.8\pm1.1$ & $3.5\pm1.0$ & $2.5\pm0.7$ & $3.3\pm0.7$ & $3.3\pm0.7$ & $0.9^{+1.5}_{-0.9}$ & $<2.2$ \\
				 
				  $2.4< \log(\rm SFR/ M_\odot\,yr^{-1}) \leq3.8$\tablenotemark{a} & $0.9\pm0.4$ & $1.5\pm0.8$ & $1.2\pm0.6$ & $2.5\pm0.8$ & $1.8\pm0.5$ & $0.9^{+0.5}_{-0.9}$ & $< 1.7$\\[0.5ex]
				  
				\hline
				  $0.1<z\leq0.5$\tablenotemark{a} & $4.2\pm1.3$ & $4.0\pm1.3$ & $1.7\pm0.5$ & $4.2\pm1.1$ & $3.8\pm1.0$ & $1.4^{+0.6}_{-0.6}$ & $1.1^{+1.2}_{-0.3}$\\
				  
				 $0.5<z\leq1.0$\tablenotemark{a} & $3.8\pm0.9$ & $3.8\pm0.9$ & $2.3\pm0.5$ & $3.6\pm0.7$ & $3.5\pm0.7$ & $1.3^{+0.9}_{-0.6}$ & $<1.8$\\
				 
				  $1.0<z\leq2.0$\tablenotemark{a} & $3.6\pm1.3$ & $3.3\pm1.2$ & $2.6\pm0.9$ & $3.4\pm0.7$ & $3.4\pm0.7$ & $1.1^{+1.8}_{-1.1}$ & $<2.2$\\
				  
				  $2.0<z\leq4.5$\tablenotemark{a} & $2.1\pm1.2$ & $1.7\pm0.9$ & $1.2\pm0.7$ & $2.0\pm0.6$ & $2.0\pm0.6$ & $ 1.0^{+0.5}_{-1.0} $ & $<1.7$ \\[0.5ex]
				  
				\hline
			\end{tabular}\\}
			\tablenotetext{a}{SFGs with $\log(M_\star/\rm M_\odot)>10$. The quoted errors refer to the  error of the median.}
	\label{table_mediansizes}
\end{table*}

Comparing the median radio, UV, and optical  size of SFGs, we find that the 3\,GHz radio emission of galaxies with $\log( M_\star/\rm M_\odot)>10$ is a factor $3.0\pm1.0$ less extended than their UV/optical emission. Note that this trend remains even if we exclude SFGs that are unreliably resolved at 3\,GHz (left panel of Figure\,\ref{fig:size-mass}).  This result is  consistent with the offset between the UV-to-optical and radio size of galaxies that has been reported in past studies \citep{murphy17, bondi18, jimenezandrade19}. Likewise, 
dust continuum  emission, tracing star formation, tends to be more compact than the  UV/optical light distribution of $\log(M_\star/\rm M_\odot)\gtrsim 10$  SFGs at $z\approx2$  \citep[e.g.,][]{simpson15,  rujopakarn16, hodge16, elbaz18, gullberg19, lang19}. Those galaxies  have a median effective radius in the FIR of $\sim1.5$\,kpc, which  is comparable with the 3\,GHz radio size of  galaxies with $\log(M_\star/\rm M_\odot)\gtrsim 10$ reported here ($\approx1.3$\,kpc).

{  We now explore the effect of our selection function (\S \ref{subsub:selection_function}) on the size -- stellar mass relation.  This is done by deriving the SFR for a typical (main sequence) galaxy at a given stellar mass and redshift. We then convert that SFR to flux density and associate it with the maximum detectable angular size presented in Figure\,\ref{fig:selection_function}.  As shown in Figure\,\ref{fig:size-mass},  \color{black} out to $z\approx0.9$, our radio selection function allows us to detect typical (main sequence) SFGs  with $\log(M_\star/\rm M_\odot)>10$ whose radio emission can be as extended as that from the UV/optical. Thus, the lack of extended radio sources is not a result of our selection function. This suggests that the centrally enhanced radio emission (relative to the UV/optical one) is a common property of the general population of massive SFGs out to $z\approx0.9$. In contrast, at $z>0.9$, our radio selection biases our sample toward compact (and bright) radio sources,  which might not be representative of the general population of massive SFGs at high redshifts. If  there exist radio sources as extended as the UV/optical size of typical (main sequence) galaxies at $z>0.9$, they will not be detected in our radio maps. Deeper radio imaging is needed to verify if the radio sizes  of the whole population of massive SFGs at $z>0.9$ (and not only radio-selected SFGs) are on average smaller than UV/optical sizes.

Lastly,} because our sample is dominated by massive SFGs that likely host a centrally concentrated dust distribution, we have to infer the effect of dust attenuation on the UV/optical size -- stellar mass relation reported above. 
  Using artificial galaxy images derived from radiative transfer simulations, it has been  shown that the observed effective radius  in the UV/optical regime is affected by dust attenuation   \citep{mollenhoff06, pastrav13}. Due to an enhanced dust content in the center of galaxies, the UV/optical emission would appear less centrally concentrated, which artificially boosts the UV/optical half-light radius. 
 Under the assumption that galaxies in our sample are randomly oriented with an average disk inclination of $i=45\degr$ and are optically thick \citep[$\tau=1$; ][see their Figure\,18]{pastrav13}, we expect that the median UV/optical sizes reported here are overestimated by $\lesssim 10\%$. Nevertheless, because more massive galaxies are more dust attenuated \citep{pannella15, nelson16},  the ratio of the apparent to the intrinsic radius of the most massive galaxies can be a factor of $\approx 2$ \citep[assuming the highest central optical depth used by][]{mollenhoff06, pastrav13}.  
 Dust attenuation could hence partially drive (a)  the positive correlation between the UV/optical size and stellar mass of galaxies in our sample, and (b) the large discrepancy (factor of $\approx3$) between the UV-to-optical  and radio size  of $\log(M_\star/ \rm M_\odot)\gtrsim10$ SFGs.

 \subsection{Size {\it vs} SFR}
 \label{subsec:size-starformation}

\begin{figure*}
	\begin{centering}
		\includegraphics[width=18cm]{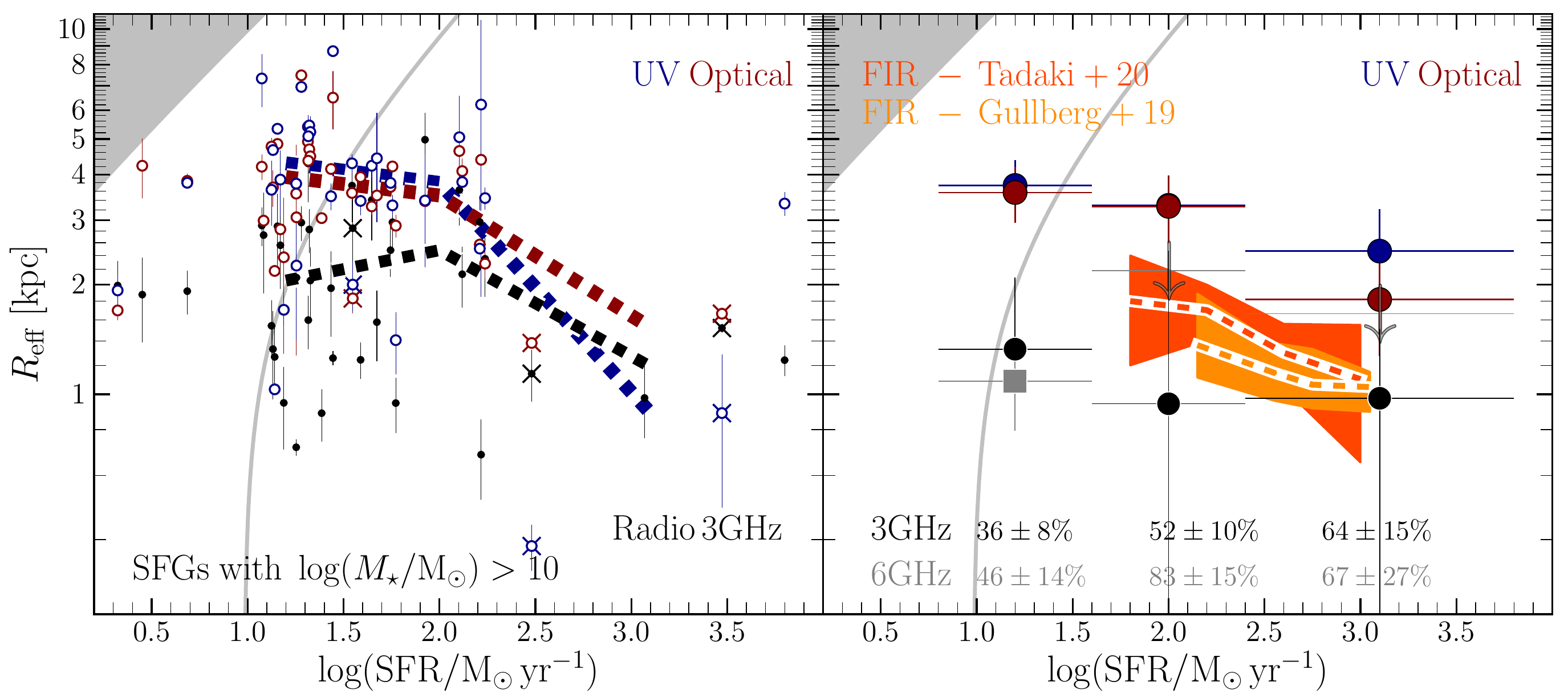}	
		\caption{{\it Left panel--} The radio/UV/optical effective radius of $\log(M_\star/\rm M_\odot)>10$ SFGs that are reliably resolved at 3\,GHz as a function of their SFR. The dashed lines show the median effective radius per SFR bin ($1-1.5$\,dex width). {   \color{black} The cross symbols denote the radio sources with a reported X-ray counterpart.} {\it Right panel--} The median radio/UV/optical effective radius  of all galaxies in the sample (i.e., sources that are reliably and unreliably resolved at 3\,GHz) per SFR bin.   The dashed lines show the  FIR size -- SFR relation of 85 massive SFGs at $1.9<z<2.6$ \citep[dark orange;][]{tadaki20} and 163 massive SFGs at $2.5<z<3.5$ \citep[orange;][]{gullberg19}. To account for the upper limits to the radio sizes of unreliably resolved galaxies,  we derive the median effective radius per stellar mass bin (and its associated uncertainty) via the KM  estimator \citep{kaplan58}. The length of the error bars indicates the error of the median and the stellar mass bin, respectively. The down arrows indicate the upper limit to the median size of galaxies per  bin. {   The grey curve shows the maximum $R_{\rm eff}$ that can be observed for a  galaxy at $z=0.9$, which is the median redshift of our galaxy sample. The grey region illustrates the parameter space that can not be probed with our radio data set.}
		The values at the bottom indicate the fraction of unreliably resolved galaxies at 3 and 6\,GHz, respectively.  In both panels,  the radio/UV/optical size of galaxies tend to decrease with SFR. The larger fraction of unreliably resolved galaxies (at both 3 and 6\,GHz) with high SFR hints toward the more compact nature of starbursts.   }
		\label{fig:size-sfr}
	\end{centering}
\end{figure*}

 To address how the galaxy size  scales with the level of star formation activity, in Figure\,\ref{fig:size-sfr} we plot the radio/UV/optical effective radius as a function of the SFR galaxies. We focus on the $\log(M_\star/\rm M_\odot)>10$ regime to mitigate the intrinsic size -- stellar mass dependence of SFGs (\S \ref{subsec:size-mass}), and to  hinder the effects of incompleteness. 
 
 We find that the 3\,GHz radio size of galaxies with SFR\,$\lesssim \rm 100\,M_\odot\, yr^{-1}$ is $1.5^{+1.5}_{-0.5}$ times larger than galaxies with the highest levels of star formation (see Table \ref{table_mediansizes}). This trend remains even if only reliably resolved sources are included in the analysis (left panel of Figure\,\ref{fig:size-sfr}), or if SFGs at $z>1$ or $z<1$ are considered. Besides, the large fraction of unreliably resolved sources at both 3\, and 6\,GHz in our highest SFR bin ($\geq 65\%$) also hints toward a compact nature of starbursts. This supports previous studies reporting that the size of galaxies measured from radio  \citep[e.g.,][]{condon91, murphy13,jimenezandrade19} and dust continuum emission tends to  decrease with increasing SFR  \citep[e.g.,][]{simpson15, rujopakarn16, gullberg19, thomson19}. Using the FIR sizes and SFRs of 85 massive SFGs at $1.9<z<2.6$ from \citet{tadaki20} and 163 massive SFGs at $2.5<z<3.5$ from \citet{gullberg19}, we  present in Figure\,\ref{fig:size-mass} the FIR size -- SFR relation. Note that across all the redshift ranges probed by \citet{gullberg19},  \citet{tadaki18}, and this study,  there appears to be a consistent trend: the  radio, FIR, and UV/optical emission tends to be more compact in SFGs with high SFRs.  We acknowledge that despite these results meet expectations from previous studies, the radio size -- SFR relation reported here is subject to a high degree of uncertainty. Further verification of this result will require a better sampling of the radio size -- SFR relation per redshift bin, which, due to our limited sample of galaxies, can not be addressed here.  
  
 {  To investigate the effect of our radio selection function on the radio size -- SFR relation, we infer the maximum recovered angular size as a function of SFR (Figure\,\ref{fig:size-sfr}). We first derive the SFR of a typical (main sequence) galaxy at a given redshift and convert it into flux density, which is subsequently related with the maximum $R_{\rm eff}$ presented in Figure\,\ref{fig:selection_function}. We find that our selection function hinders the identification of extended galaxies with low SFRs. Thus, recovering this underrepresented galaxy population  will further support the anti-correlation between the radio size and SFR of galaxies.}
 
 In Figure\,\ref{fig:size-sfr}, we also present the UV/optical size -- SFR relation for SFGs that are reliably resolved at 3\,GHz (left panel) and all SFGs in the sample (i.e., reliably and unreliably resolved; right panel). In both cases, the UV/optical size of galaxies tends to decrease with increasing SFR: galaxies with $\rm SFR\lesssim \rm 10\,M_\odot\, yr^{-1}$ are  $2.0\pm0.5$ times more extended than galaxies producing stars at a higher rate (i.e., $\rm SFR\gtrsim\rm 100\,M_\odot\, yr^{-1}$, Table \ref{table_mediansizes}).  
 These results are consistent with \citet{elbaz11} and \citet{wuyts11} who derive rest-frame UV and optical sizes, respectively, and show that starbursts  are more compact than  MS galaxies over the redshift range $0 \lesssim z \lesssim 2$. 
 
Despite the apparent trend between the UV/optical size and SFR, we still have to consider whether this result is driven by dust extinction. As we discussed in \S \ref{subsec:wave-vs-hstsize},  dust can dim the UV emission from the galaxy's central region  and, as a result, the apparent UV effective radius is artificially enlarged. Since  starbursts host a larger dust content than MS galaxies \citep{liu19b}, we expect that the (observed) UV effective radius of starbursts  tends to be   overestimated by a larger fraction. Correcting for this effect  would further strengthen the contrast between the UV size of galaxies with low ($\lesssim \rm 10\,M_\odot\, yr^{-1}$) and high SFR ($\gtrsim\rm 100\,M_\odot\, yr^{-1}$).

{  Although our radio-selected sample is expected to be dominated by SFGs (\S \ref{subsub:agn_fraction}), another potential bias affecting the radio size -- SFR relation is the contribution from   \color{black} a point-like radio component from an AGN. To address this issue, we exclude the eight AGN candidates (\S \ref{subsub:agn_fraction}) falling within the SFR bins  presented in Figure\,\ref{fig:size-sfr} (left panel). These AGN candidates represent $45\pm12\%$ of all galaxies in our highest SFR bin, which contrasts with the low fraction ($\lesssim 10\%$) of AGN candidates in the intermediate and lowest SFR bin. By removing such AGN candidates from the analysis, we corroborate that the direction of the radio size -- SFR relation presented in Figure\,\ref{fig:size-sfr} is not significantly affected. On the other hand, the median UV/optical size of galaxies in the highest SFR bin becomes  uncertain after removing the AGN candidates (see Figure\,\ref{fig:size-sfr_noagn}). A more statistically significant sample, as well as  dedicated multi-wavelength observations to identify AGN-dominated systems, will be needed to verify that the UV/optical size of ``pure SFGs" decreases with increasing SFR.

Focusing on the radio size -- SFR relation}, the {  apparent} compactness of highly active SFGs  can {  also} be interpreted within the context of a  two-component model.  The first component  is a compact, dusty starburst ($R_{\rm eff}\sim 1\rm \,kpc$), and the second is a larger ($R_{\rm eff}\sim 5\rm \,kpc$), less active disk \citep[see][]{gullberg19, thomson19}.   The relative brightness of these two components hence varies as a function of the total SFR of the galaxy.  In highly active systems with  SFR\,$\gtrsim 100\,\rm M_\odot\,yr
^{-1}$, such as  Ultra-Luminous Infrared Galaxies  \citep[ULIRGS; e.g.,][]{wilson14},  the small nuclear emission dominates, whereas in more passive galaxies with lower specific SFRs (i.e., SFR/$M_\star$) the extended component becomes dominant \citep{ellison18}. The effective radius that we measure from the radio, tracing star formation, is thus a weighted sum of these two components, leading to the dependence of  $R_{\rm eff}$ on the SFR of galaxies.

The physical mechanisms driving intense star formation in the central component of galaxies have been linked to gas-rich mergers, which  channel gas into the center of galaxies and trigger central, compact starbursts \citep{mihos96, hopkins06}. This is the case of local/low-redshift ULIRGS that exhibit high levels of star formation activity \citep{genzel01}. However, at high redshifts, galaxies harbor  more massive gas reservoirs than their low-redshift counterparts \citep[e.g., ][]{liu19b, birkin20}, potentially rendering violent disk instabilities (VDI)  more common and efficient in driving gas inflows \citep[e.g.,][]{bournaud07,  bournaud09,   dekelburkert14}. This may lead to  an abrupt enhancement of  the galaxy's central cold gas  surface density,  triggering  compact starbursts in high-redshift galaxies  \citep{fensch17, wang19}.

 \subsection{Size {\it vs} redshift}\label{subsec:size-redshift}

\begin{figure*}
	\begin{centering}
		\includegraphics[width=18cm]{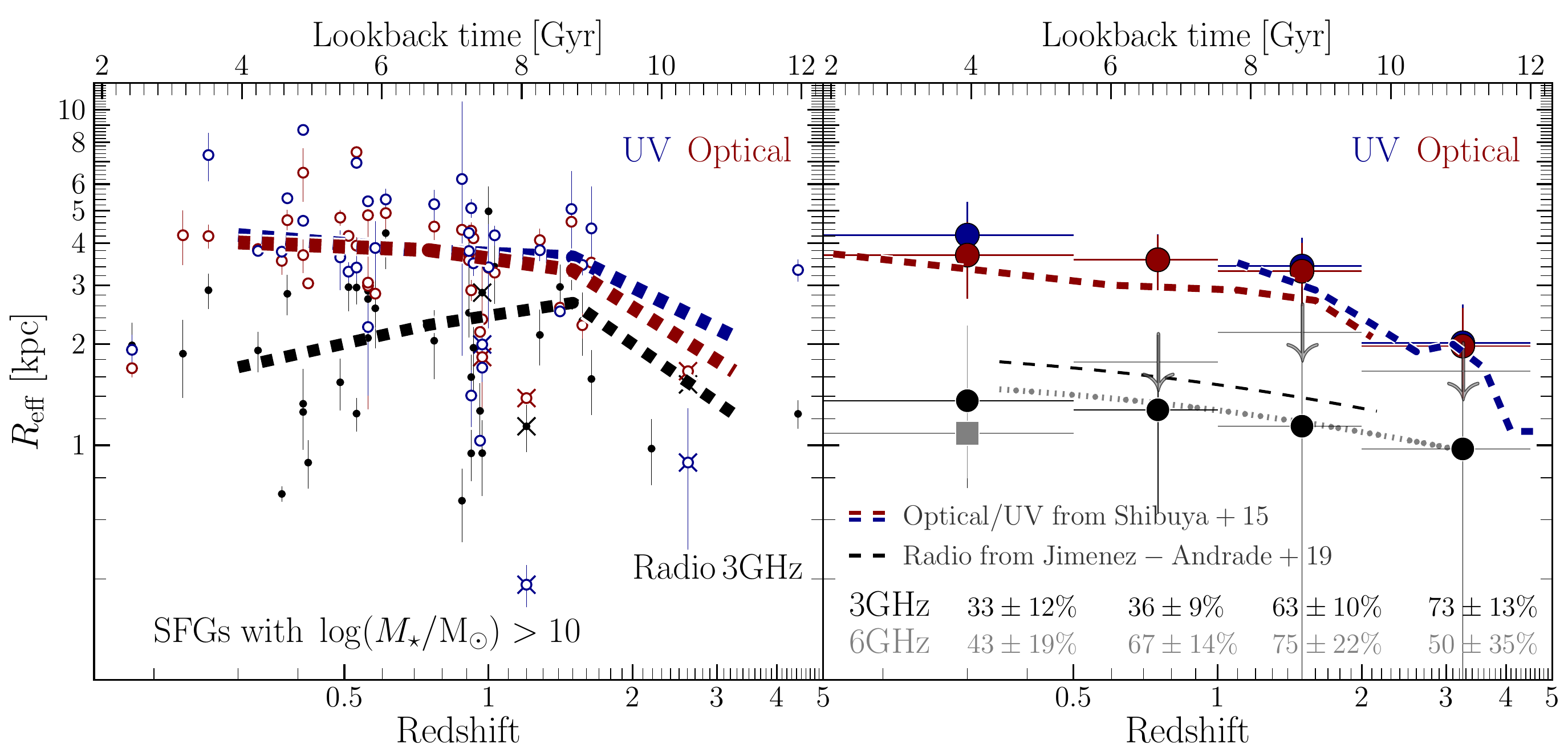}	
		\caption{
		{\it Left panel--} Redshift evolution of the radio/UV/optical effective radius  of SFGs with $\log(M_\star/\rm M_\odot)>10$  that are reliably resolved at 3\,GHz. The dashed lines show the median effective radius per redshift bin ($\approx 2$\,Gyr width). {   \color{black} The cross symbols denote the radio sources with a reported X-ray counterpart.} {\it Right panel--} The median radio/UV/optical effective radius of all galaxies in the sample (i.e., sources that are reliably and unreliably resolved at 3\,GHz) per redshift bin. The values at the bottom indicate the fraction of unreliably resolved galaxies at 3 and 6\,GHz, respectively. The solid black line shows the redshift evolution of the UV/optical size of SFGs with $\log(M_\star/{\rm M_\odot})=10.5-11$ derived by \citet{shibuya15}, which is the stellar mass range that resembles the median size of SFGs in our sample ($\log(M_\star/{\rm M_\odot})=10.6$). The dashed black line shows the redshift evolution of the 3\,GHz radio size of ($\log(M_\star/{\rm M_\odot})> 10.5$ SFGs reported by \citet{jimenezandrade19}. The dotted blue line is a  fit to the 3\,GHz data points presented here, leading to $R_{\rm eff}=(1.6\pm 0.4) (1+z)^{-0.3\pm 0.3}$\,kpc. This plot demonstrates that the UV/optical size of massive SFGs is significantly more extended than the 3\,GHz radio size out to $z\approx3$,  and that the UV/opitcal/radio size of  SFGs with $\log(M_\star/\rm M_\odot)>10$ evolve with redshift at a similar rate.}   
		\label{fig:size-redshift}
	\end{centering}
\end{figure*}

To explore the growth rate of galaxies across cosmic time, we first evaluate our selection function to account for  potential biases in deriving the radio/UV/optical size evolution of galaxies. A well-known observational bias is that the most distant galaxies of flux-limited samples tend to be bright sources. As a result, the faint galaxy population is systematically underrepresented at high redshifts. 
Our radio detection limit of $\approx 4.5\,\rm \mu Jy$  certainly imposes an SFR threshold above which we can detect galaxies of a given SFR and redshift (see Figure\,\ref{fig:sample_properties}). Converting this flux density limit into SFR and using  the parametrization of the MS from \citet{schreiber15}, we verify that selecting galaxies in our sample with $\log(M_\star/\rm M_\odot)>10$ (10.5)  allow us to probe 
``typical'', main sequence SFGs out  to $z\approx2$ ($z\approx3$) (see Figure\,\ref{fig:sample_properties}). Hence, by adopting a  mass-selection limit of $\log(M_\star/\rm M_\odot)\approx10$, we are sensitive to both typical and highly active SFGs (starbursts) out to $z\approx2$. At higher redshifts, $z\gtrsim2$, our sample is biased toward the starburst population. Although here we present the  size evolution of SFGs out to $z\approx3$ (Figure\,\ref{fig:size-redshift}), we acknowledge that the information from our $z=3$ bin is affected by incompleteness, and hence the conclusions drawn from it must be interpreted with caution.

 In Figure\,\ref{fig:size-redshift}, we present the size evolution of SFGs  with $\log(M_\star/\rm M_\odot)>10$. In the left panel, we only present SFGs that are reliably resolved at 3\,GHz. In the right panel, we show SFGs that are reliably and unreliably resolved at 3\,GHz. In both cases, the  UV/optical size of SFGs increases by a factor of $2.0\pm 0.4$ from  $z\approx3$ to $z\approx0.3$ (Table \ref{table_mediansizes}, resembling the trend reported by \citet{shibuya15} for SFGs with a consistent stellar mass (see right panel of Figure\,\ref{fig:size-redshift}). 
 On the contrary, the radio size -- redshift relation behaves differently if only SFGs that are reliably resolved at 3\,GHz are considered. This is a consequence of the increasing fraction of unresolved galaxies at higher redshifts (right panel of Figure\,\ref{fig:size-redshift}). Including only reliably resolved sources in the analysis misses the population of high-redshift, compact radio sources that are not yet resolved in our  radio images. We, therefore, rely on the median radio sizes derived via the KM estimator that allow us to consider both reliably and unreliably resolved radio sources (see Table \ref{table_mediansizes}).  This analysis indicates that the  radio size evolution of  SFGs in our sample  resembles that of massive SFGs reported by \citet{bondi18} and \citet{jimenezandrade19} (see right panel of Figure\,\ref{fig:size-redshift}). 
 Qualitatively, we find that the 3\,GHz radio size of SFGs evolves with redshift as  $R_{\rm eff} = (1.6\pm0.4) (1+z)^{-0.3\pm 0.3}$\,kpc out to $z\approx3$, which is in agreement with  \citet{jimenezandrade19}\footnote{\citet{jimenezandrade19} reports the size evolution for MS and starburst galaxies separately. Here, we use their public catalog to derive the size evolution of the total SFG population (i.e., MS and starburst galaxies), to allow for a more consistent comparison with our results.} who report $R_{\rm eff} = (2.0 \pm 0.2)(1+z)^{-0.4\pm 0.1}$\,kpc out to $z=2.35$ (see Figure\,\ref{fig:size-redshift}). The $\approx$20\% lower median 3\,GHz size of galaxies in our sample than the one reported by \citet{bondi18} and \citet{jimenezandrade19} is likely a consequence of different selection criteria, as the latter studies consider more massive SFGs with $\log(M_\star/\rm M_\odot)\gtrsim 10.5$ over $0.35 < z \lesssim 3$.  {   \color{black}  Lastly, we  verify that the radio/UV/optical size evolution of galaxies presented in Figure\,\ref{fig:size-redshift} is not significantly affected if we exclude the 12 AGN candidates in our sample (see Figure\,\ref{fig:size-redshift_noagn} in the Appendix).}

The radio continuum size evolution of galaxies has been explored at different frequencies.  \citet{lindroos18} find that the 1.4\,GHz radio size of SFGs also decreases with redshift as  $R_{\rm eff} \approx 6(1+z)^{-1.7}$\,kpc (see their Figure\,3). This indicates that the growth rate and radio size at 1.4\,GHz are   higher than those from the 3\,GHz radio emission, possibly due to the frequency-dependent cosmic ray diffusion that leads to more extended radio sizes at lower frequencies (see \S \ref{subsec:cosmicrays}). Indeed,  the $\approx$1.4\,GHz  continuum emission can be as extended as the stellar light distribution of $1\lesssim z \lesssim 3$ SFGs \citep{muxlow05, owen18, muxlow20}, while the 10\,GHz radio size of galaxies at similar redshifts is always smaller (or equal) than the optical size \citep{murphy17}.

The emerging consensus, supported by our  observations, is that the $\gtrsim$\,3\,GHz radio size of most SFGs remains a factor of three  more compact than their optical size out to $z\approx 3$ \citep{murphy17,  bondi18,   jimenezandrade18}. 
Because dust attenuation alone does not seem to account for the large discrepancy between the radio and optical emission of SFGs   (see \S \ref{subsec:size-mass}), our results from Figure\,\ref{fig:size-redshift} suggest that star formation  --traced by $\nu \gtrsim3$\,GHz radio continuum imaging -- remains centrally concentrated in galaxies across cosmic time. Obtaining reliably dust-corrected UV size of galaxies will be crucial to verify this result \citep{lang19}, as it should also trace the centrally enhanced, massive star formation activity revealed by the radio continuum emission.

\subsection{The effect of differential cosmic-ray electron diffusion on  radio size measurements}
\label{subsec:cosmicrays}
 
  Mapping non-thermal (synchrotron) radio emission accelerated by SNRs allows us to trace the sites of massive star formation across galactic disks. However, CREs arising from these SNR can  propagate further through the ISM \citep[e.g., ][]{murphy08, murphy09, murphy12b, berkhuijsen13, kobayashi04},  biasing to some extent the spatial distribution of star formation in galaxies as traced by radio emission. Because the diffusion length ($l_{\rm diff}$) depends on the CRE emitting frequency  and the physical properties of the ISM \citep[e.g., ][]{murphy08, murphy09}, here we discuss how CRE diffusion affects the radio size and its dependence on redshift and star formation. 
  
   As the redshift increases, our 3\,GHz radio imaging probes higher rest-frame frequencies given by $\nu_{\rm em}=3{\rm GHz} (1+z)$. Due to the rapid cooling time (i.e., short lifetime) of high-frequency emitting CRE, these cannot propagate as far as those at lower frequencies. Thus, the more compact radio size of high-redshift galaxies (see Figure\,\ref{fig:size-redshift}) might be a consequence of the systematically shorter diffusion length of CRE emitting at $\nu>3{\rm GHz}$. 
      Using the formulae in \citet[][see their Section 7.1]{murphy08}, as well as additional considerations for radiative losses, escape and cosmic microwave background (CMB) effects included in \citet{murphy09}, we  estimate the diffusion length of CRE as a function of frequency. This is done by assuming that the propagation of CRE is  described by a simple random walk process, the diffusion length can be approximated as $l_{\rm diff}= (\tau_{\rm diff} D_{  R})^{1/2}$ \citep[e.g., ][]{murphy08, murphy09, murphy12b}; where  $D_{ R}$ is the  rigidity ($R$)-dependent spatial diffusion coefficient and $\tau_{\rm diff}$ is the CRE travel time. $D_{ R}=D_0 [R/{\rm GV}]^\delta=D_0 [(E_{\rm CR}^{2}-E_0^2)^{1/2}/\rm GV]^\delta$, where $E_{\rm CR}$  is the CRE energy,  $E_0$ is the particle rest-mass energy,  $D_0$ is the normalization constant,  
      and $\delta\sim 0.7$ \citep{murphy12b}. Lastly, $\tau_{\rm diff}$ is estimated following \citet{murphy09}, which takes into account both radiative losses and escape of CREs, along with CMB effects.

      Based on these estimates, we find that, at $z\approx0.3$, the observed 3\,GHz radio emission corresponds to CRE emitting at $\nu=4$\,GHz,  which can diffuse  $\approx 1.6$ times further through the ISM than CRE emitting at $\nu=12$\,GHz at $z\approx 3$. This  observational bias, therefore, could partially account for the $\approx 2$ times larger radio size of galaxies at $z\approx 0.3$ than their high-redshift counterparts at $z\approx3$. To mitigate the effect of differential CRE diffusion on the radio size evolution of galaxies,  we require high-frequency radio observations that probe the thermal (free-free) radiation from massive star-forming regions \citep[e.g., at $\gtrsim$10\,GHz; see][]{murphy17}.   
      
       The age of the CRE populations can also affect the surface brightness distribution, and hence the effective radius, of non-thermal synchrotron emission of galaxies \citep{murphy08}. In the case of starbursts, the CRE population is dominated by young, freshly injected particles that have yet to propagate significantly from their birthplaces.  Consequently, the non-thermal radio continuum emission can appear more concentrated in starburst galaxies. This physical phenomenon could explain the tentative evidence for more compact radio sizes with increasing  star formation activity (see \S \ref{subsec:size-starformation}), as also reported in  detail by \citet{jimenezandrade19}. Furthermore, the UV/optical emission also appears more centrally concentrated in starburst galaxies (see Figure\,\ref{fig:size-sfr}), also suggesting that the compact radio size of the starbursts reflects a high central SFR surface density that is likely dominated by a population of young, freshly accelerated CREs that yet to propagate significantly into the galaxy disks. Under this scenario, the smaller radio sizes at high redshifts  could  reflect  the increasing  contribution from a nuclear starburst relative to a less active, extended disk-like component \citep{thomson19}. Verifying this hypothesis will require even deeper and higher resolution radio maps to decompose the surface brightness profile into compact and disk components, as currently done with ALMA observations of the dust continuum with a $\rm FWHM \lesssim0\farcs2$ resolution \citep[e.g.,][]{gullberg19,hodge19}.  {  Decomposing the radio emission profile of high-redshift SFGs will also allow us to trace the assembly of the extended stellar body of the massive SFGs at $z<3$ in our sample, as we expect that these structures might be the remnant of widespread --rather than centrally enhanced-- star formation activity across $z>3$ disks. }\\
 
\section{Summary}
To better understand the mechanisms driving the stellar mass buildup in star-forming galaxies,  we use VLA (Heywood et al.\,{  in press}) and  {\it HST} ACS/WFC3 \citep{shipley18}  imaging  to measure and compare the rest-frame radio and UV/optical size of 98 star-forming galaxies in the {\it Hubble} Frontier Fields. While radio continuum radiation probes the bulk of the  massive star formation in galaxies, the optical emission traces the stellar disk of galaxies. Our radio-selected sample comprises 98 star-forming galaxies  over $0.3\lesssim z 
\lesssim 3$, with a median redshift of $z\approx0.9$ and median stellar mass of $\log(M_\star/\rm M_\odot)\approx 10.4$.  Our main results are the following:

\begin{itemize}
    \item  The median 3\,GHz radio size for all the 98 galaxies in our sample is $R_{\rm eff}=1.3\pm0.3$\,kpc. Among these, 31 have 6\,GHz counterparts. Their median 3 and 6\,GHz radio effective radii are $1.3\pm0.3$ and  $1.1^{+0.7}_{-0.3}$\,kpc, respectively. This implies a ratio of 3-to-6\,GHz radio size of  $1.1\pm0.1$. 
    
    \item The UV/optical size of galaxies increases with the stellar mass (as widely reported in the literature). In contrast, there is no clear dependence between the 3\,GHz radio size and  stellar mass of SFGs. Thus, because more massive galaxies are more heavily dust-obscured, it is likely that the UV/optical size -- stellar mass relation is partially driven by dust extinction effects.  
    
    \item The radio  size of massive galaxies with  $\log(M_\star/\rm M_\odot)> 10$  decreases with increasing SFR. SFGs with SFR\,$\lesssim10\,\rm M_\odot \, yr^{-1}$ are on average $1.5-2.0$ times more extended than the most active systems with  SFR\,$\gtrsim100\,\rm M_\odot \, yr^{-1}$. {   \color{black} Removing AGN candidates from the analysis does not significantly affect this result. The median UV/optical size of galaxies with  $\log(M_\star/\rm M_\odot)> 10$ appears to follow a similar trend, yet it remains unclear if this relation is driven by a higher incidence of AGN  in galaxies with SFR\,$\gtrsim100\,\rm M_\odot \, yr^{-1}$.}
    
    \item   The 3\,GHz radio size increases with cosmic time as $R_{\rm eff}/{\rm kpc}\propto (1+z)^{-0.3\pm0.3}$ across $z\approx3$ to $z\approx0.3$. Similarly, the UV/optical size of massive SFGs with  $\log(M_\star/\rm M_\odot)> 10$ increases by a factor of $2.0\pm0.4$ from $z\approx3$ to $z\approx 0.3$. Over the cosmic epoch probed here,  the radio size of most SFGs in our {  radio-selected} sample remains a factor of two-to-three more compact than their UV/optical size. {   \color{black} This trend appears to be present in the general population of massive SFGs over $0.1\lesssim z\lesssim 1$, which is the redshift range that is not significantly affected by our radio selection function. }
    
 \end{itemize}

Overall, our results indicate that massive, {  radio-selected} SFGs show centrally enhanced star formation activity relative to their outskirts, possibly due to {  a large concentration} of cold gas in the galaxy center generated by VDI- and/or merger-driven gas inflows.  The smaller sizes of the most active, {  radio-selected} SFGs suggests that star formation in these galaxies mainly occurs within a central, compact starburst, while less active systems harbor more widespread star formation across the galaxy's disk.  Verifying  these trends requires higher-resolution radio observations (FWHM $\lesssim 0\farcs6$) to decompose the surface brightness profile of radio continuum emission of SFGs into bulge/disk  components, {  as well as dedicated multi-wavelength observations to disentangle the contribution of AGN to the light profile of their host galaxies}. Higher frequency radio observations will be paramount as well, as $\gtrsim 10$\,GHz radio emission probes the thermal (free-free) radiation that is  a better tracer of (instantaneous) massive star formation of $z \gtrsim 1$ galaxies \citep{murphy17}.

\acknowledgments
{  We thank the reviewer for their careful reading of the manuscript and their constructive comments. }
EFJA gratefully acknowledges the support from the NRAO staff that made remote working feasible during the COVID-19 pandemic. IH acknowledges support from the UK Science and Technology Facilities Council (ST/N000919/1); the Oxford Hintze Centre for Astrophysical Surveys which is funded through generous support from the Hintze Family Charitable Foundation; and the South African Radio Astronomy Observatory which is a facility of the National Research Foundation (NRF), an agency of the Department of Science and Innovation. IS acknowledges support from STFC (ST/T000244/1).     
Based on observations made with the NASA/ESA {\it Hubble Space Telescope}, obtained from the data archive at the Space Telescope Science Institute. STScI is operated by the Association of Universities for Research in Astronomy, Inc. under NASA contract NAS 5-26555. Part of this research was carried out at the Jet Propulsion Laboratory, California Institute of Technology, under a contract with the National Aeronautics and Space Administration. Support for this work was provided by NASA through grant number HST-AR-14306.001-A from the Space Telescope Science Institute, which is operated by AURA, Inc., under NASA contract NAS 5-26555. The National Radio Astronomy Observatory is a facility of the National Science Foundation operated under cooperative agreement by Associated Universities, Inc.

%

\facilities{{\it HST} (ACS, WFC3), VLA} 


\software{ {\tt Astropy} \citep{astropy13, astropy18},  {\sc eazy} \citep{brammer08}, {\sc fast} \citep{kriek09}, {\tt lifelines} \citep{cameron_davidson_pilon_2020_4136578}, {\sc PyBDSF} \citep{Mohan15}, {\tt SciPy} \citep{scipy20},   {\tt statmorph} \citep{rodriguezgomez18}.}



\clearpage

\appendix
\section{Comparing star formation rate tracers}
\label{subsec:sfr_tracers}

\begin{figure*}[h!]
 	\begin{centering}
 		\includegraphics[width=7.cm]{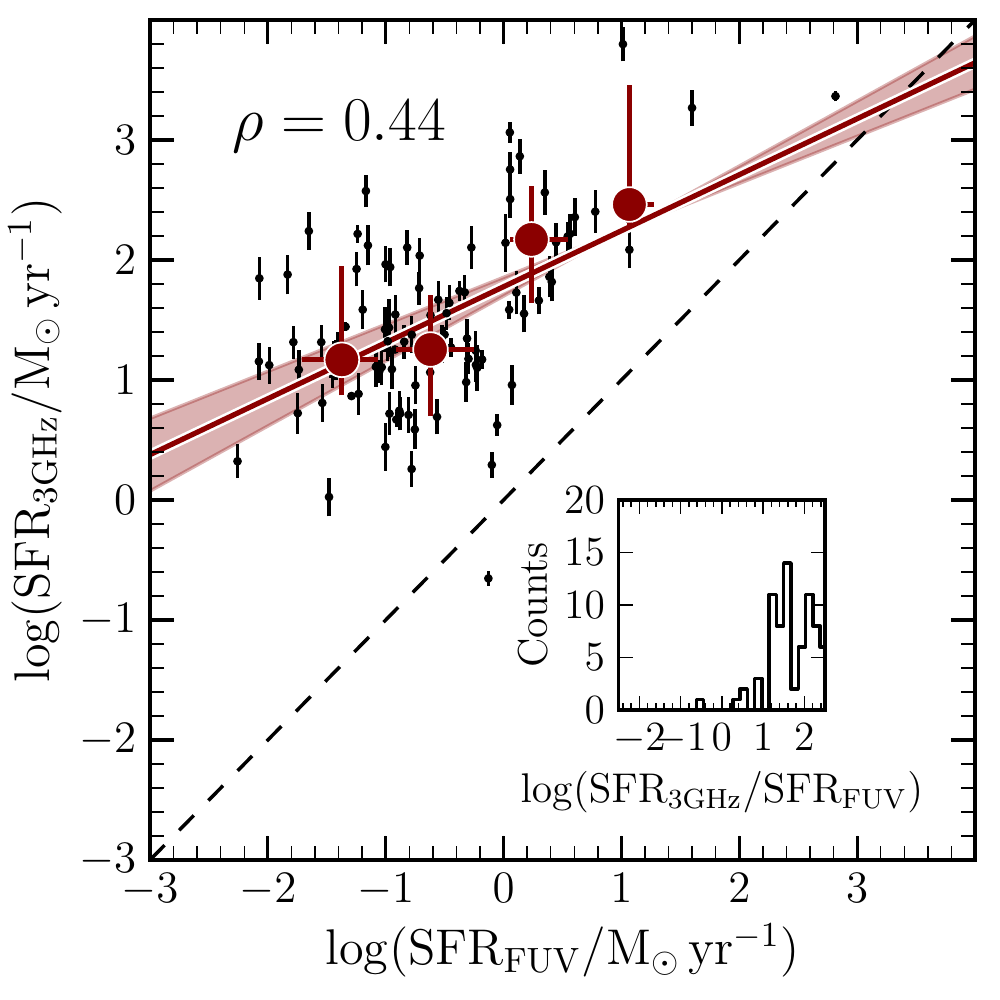}	
 		\includegraphics[width=7.cm]{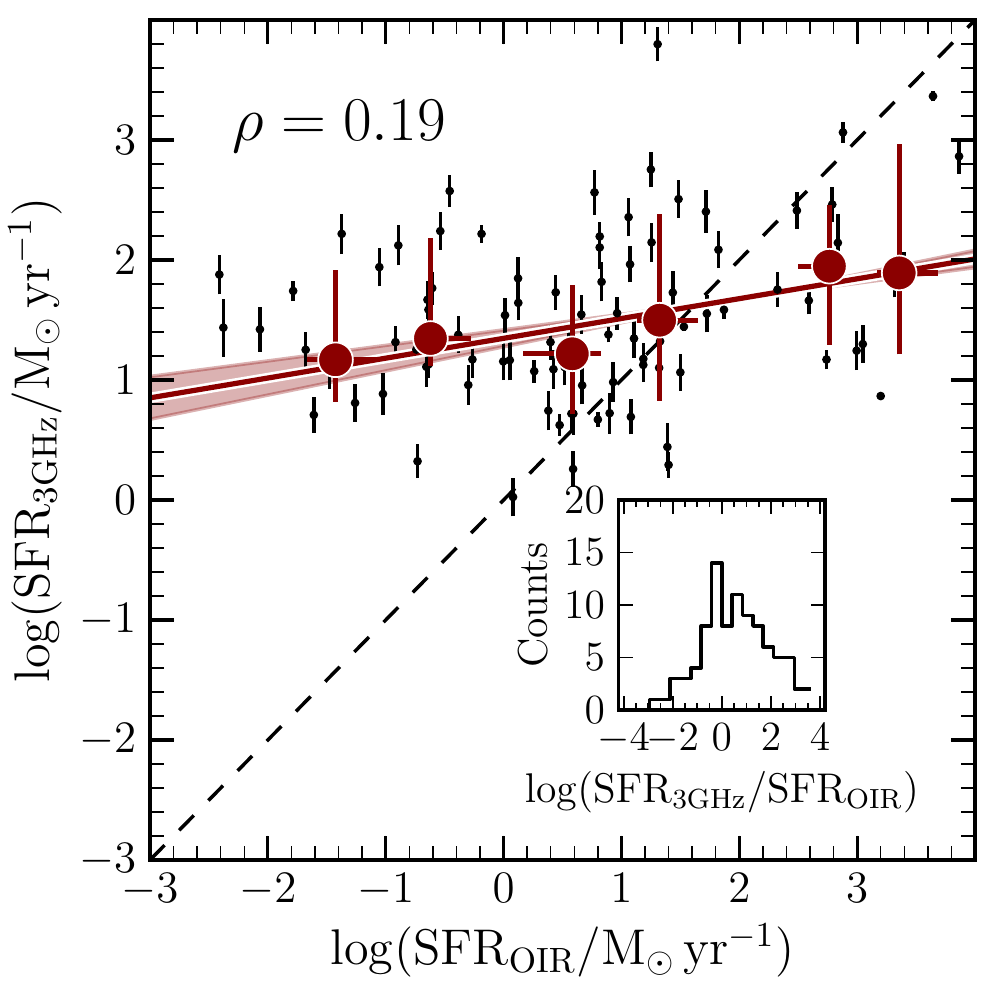}	
 		\caption{Comparison of the SFR estimated from the 3\,GHz radio flux density ($\rm SFR_{3\,GHz}$) with that inferred from the FUV ($\rm SFR_{FUV}$, {\it left panel}) and OIR ($\rm SFR_{OIR}$, {\it right panel}). 	The red circles show the median value per SFR bin, while the length of the bars corresponds to their respective 16th/84th percentiles.   The dashed black line illustrates the one-to-one relation. The solid red line is the best linear fit to the data points (black circles), whereas the shaded blue region illustrates the uncertainty of such a fit.  The inset image shows a histogram of the ordinate to abscissa ratio. The  Spearman correlation coefficient, $\rho$, is presented at the upper-left corner of each panel.   }
 		\label{fig:sfr_tracers}
 	\end{centering}
 \end{figure*}

In this work, we use radio and FUV imaging to trace the dust-obscured and unobscured star formation activity of galaxies in the sample (see \S \ref{subsec:deriving_sfr}). This information complements existing SFR estimates in the HFF DeepSpace catalog derived via optical-to-near IR (OIR) SED fitting with the {\sc fast} code \citep{shipley18}. Comparing all these distinct SFR tracers is, therefore, relevant to provide scaling relations that allow one to infer  extinction-free SFR estimates when only FUV- or OIR-based SFR are available. In Figure\,\ref{fig:sfr_tracers}, we compare our  extinction-free radio SFR estimates ($\rm SFR_{3\,GHz}$)   with the dust-biased SFR indicators from the FUV-to-OIR regime   ($\rm SFR_{FUV}$ and $\rm SFR_{OIR}$). Fitting a linear function to the  $\rm SFR_{3\,GHz}-SFR_{FUV}$ relation (in log-log space), we find that  
  \begin{equation}
  \begin{split}
   \rm \log\left( \frac{SFR_{3\,GHz}}{M_\odot\,yr^{-1}} \right)=  \rm  (0.47\pm0.07) \log\left( \frac{SFR_{FUV}}{M_\odot\,yr^{-1}} \right)+(1.78\pm0.08), 
 \end{split} 
\end{equation}
\noindent 
indicating that  $\rm SFR_{3\,GHz}$ and $\rm SFR_{FUV}$ are  positively correlated. Yet $\rm SFR_{3\,GHz}$ is on average $\gtrsim10$ times larger. Similarly, in the $\rm SFR_{3\,GHz}-SFR_{OIR}$ plane we find that
\begin{equation}
\begin{split}
  \rm  \log\left(  \frac{SFR_{3\,GHz}} {M_\odot\,yr^{-1}} \right) =  \rm   (0.17\pm0.04) \log\left( \frac{SFR_{OIR}}{M_\odot\,yr^{-1}} \right)+(1.34\pm0.07).
\end{split}    
\end{equation}
\noindent 
A Spearman correlation coefficient of only 0.19 and  p-value of 0.06 suggest that $\rm SFR_{3\,GHz}$ and $\rm SFR_{OIR}$ are poorly correlated, in addition to the scatter of the relation being $\approx1$\,dex. 

\section{The multi-wavelength size of galaxies in the HFF}
In Table \ref{table_sllsizes}, we present the redshift, total SFR, stellar mass, and radio/UV/optical sizes of 98 field galaxies in our sample. We also present the properties of 15 cluster galaxies that are excluded from our analysis. An online version of this table can be found at \url{https://science.nrao.edu/science/surveys/vla-ff/data}.
\begin{longrotatetable}
\begin{deluxetable*}{ l c c c c c c c c c c c c c c}
\tablewidth{700pt}
\tabletypesize{\scriptsize}
		\tablecaption{The radio/UV/optical size of 98  field (and 15 cluster) galaxies in the HFF.}
                \tablehead{ \colhead{ID}  & \colhead{$z$} &  $z_{\rm type}$\tablenotemark{a} & $\mu$ &  $\log(M_\star/\rm M_\odot)$ & $\rm \log(SFR/M_\odot\,yr^{-1})$\tablenotemark{b} & $R_{\rm eff}^{\rm UV}$/kpc   & $R_{\rm eff}^{\rm opt}$/kpc & $R_{\rm eff}^{\rm 3\,GHz}$/kpc & Res. flag\tablenotemark{c} & $R_{\rm eff}^{\rm 6\,GHz}$/kpc &  Res. flag\tablenotemark{d} & Env.\tablenotemark{e} }
               \startdata
  VLAHFF-J041602.04$-$240523.5 &   $2.19$ &    {\it p}  &  $1.66$ &   $10.14$ &   $3.07\pm0.09$ &   $\dots$ &   $\dots$ &   $0.98\pm0.22$ &    $\checkmark$ &   $1.25\pm0.47$ &      &    f \\ 
  VLAHFF-J041605.30$-$240520.6 &   $0.39$ &    {\it s}  &  $1.0$ &   $11.13$ &   $0.85\pm0.1$ &   $2.62\pm0.35$ &   $3.69\pm0.26$ &   $1.7\pm0.38$ &    $\checkmark$ &   $\dots$ &      &    c \\ 
  VLAHFF-J041606.36$-$240451.2 &   $0.74$ &    {\it s}  &  $3.03$ &   $10.36$ &   $1.11\pm0.16$ &   $5.71\pm0.46$ &   $4.6\pm0.28$ &   $0.81\pm0.36$ &      &   $0.0\pm1.08$ &      &    f \\ 
  VLAHFF-J041606.62$-$240527.8 &   $1.9$ &    {\it p}  &  $2.26$ &   $10.03$ &   $1.75\pm0.18$ &   $\dots$ &   $\dots$ &   $1.45\pm0.49$ &      &   $1.55\pm0.74$ &      &    f \\ 
  VLAHFF-J041607.67$-$240438.7 &   $0.4$ &    {\it s}  &  $1.0$ &   $12.1$ &   $0.13\pm0.03$ &   $3.43\pm0.27$ &   $6.81\pm0.21$ &   $2.45\pm0.31$ &    $\checkmark$ &   $0.71\pm0.11$ &    $\checkmark$ &    c \\ 
  VLAHFF-J041607.89$-$240623.4 &   $0.39$ &    {\it s}  &  $1.0$ &   $10.23$ &   $1.24\pm0.06$ &   $3.59\pm0.09$ &   $3.85\pm0.17$ &   $2.96\pm0.4$ &    $\checkmark$ &   $2.11\pm0.61$ &    $\checkmark$ &    c \\ 
  VLAHFF-J041608.55$-$240522.0$\,\ast$ &   $0.97$ &    {\it s}  &  $1.48$ &   $10.59$ &   $1.55\pm0.14$ &   $2.0\pm0.31$ &   $1.83\pm0.16$ &   $2.85\pm0.63$ &    $\checkmark$ &   $\dots$ &      &    f \\ 
  VLAHFF-J041609.11$-$240459.2 &   $0.71$ &    {\it s}  &  $1.77$ &   $9.79$ &   $0.73\pm0.17$ &   $1.88\pm0.75$ &   $1.46\pm0.02$ &   $0.0\pm0.84$ &      &   $\dots$ &      &    f \\ 
  VLAHFF-J041610.62$-$240407.4 &   $0.41$ &    {\it s}  &  $1.06$ &   $10.35$ &   $1.13\pm0.11$ &   $4.66\pm0.04$ &   $3.69\pm0.42$ &   $1.33\pm0.36$ &    $\checkmark$ &   $1.46\pm0.79$ &      &    f \\ 
  VLAHFF-J041610.79$-$240447.5$\,\ast$ &   $2.09$ &    {\it s}  &  $1.98$ &   $9.82$ &   $2.12\pm0.14$ &   $0.4\pm0.21$ &   $1.54\pm0.22$ &   $1.63\pm0.62$ &      &   $\dots$ &      &    f \\ 
  VLAHFF-J041611.61$-$240221.6 &   $0.8$ &    {\it p}  &  $1.16$ &   $9.69$ &   $1.26\pm0.15$ &   $3.66\pm0.05$ &   $3.53\pm0.05$ &   $3.04\pm1.14$ &      &   $\dots$ &      &    f \\ 
  VLAHFF-J041611.67$-$240419.6 &   $2.2$ &    {\it p}  &  $2.26$ &   $10.16$ &   $1.94\pm0.15$ &   $3.22\pm0.31$ &   $2.54\pm0.03$ &   $2.22\pm0.84$ &      &   $\dots$ &      &    f \\ 
  VLAHFF-J041613.23$-$240319.8 &   $0.91$ &    {\it s}  &  $1.68$ &   $10.65$ &   $1.75\pm0.08$ &   $3.79\pm0.06$ &   $3.7\pm0.02$ &   $2.48\pm0.39$ &    $\checkmark$ &   $2.65\pm0.81$ &      &    f \\ 
  VLAHFF-J041614.21$-$240359.4 &   $0.31$ &    {\it s}  &  $1.0$ &   $10.54$ &   $0.43\pm0.11$ &   $8.97\pm0.38$ &   $7.17\pm0.44$ &   $1.27\pm0.38$ &      &   $1.8\pm0.77$ &      &    f \\ 
  VLAHFF-J041627.72$-$240645.2 &   $3.27$ &    {\it p}  &  $1.02$ &   $9.96$ &   $2.57\pm0.17$ &   $2.17\pm0.1$ &   $\dots$ &   $0.0\pm1.49$ &      &   $\dots$ &      &    f \\ 
  VLAHFF-J041630.23$-$240553.0 &   $1.59$ &    {\it p}  &  $1.28$ &   $10.93$ &   $1.84\pm0.16$ &   $3.51\pm0.22$ &   $3.36\pm0.26$ &   $0.0\pm2.03$ &      &   $\dots$ &      &    f \\ 
  VLAHFF-J041630.30$-$240630.7 &   $1.72$ &    {\it p}  &  $1.13$ &   $10.29$ &   $2.12\pm0.15$ &   $2.61\pm0.47$ &   $2.62\pm0.07$ &   $0.0\pm0.46$ &      &   $\dots$ &      &    f \\ 
  VLAHFF-J041636.19$-$240759.7$\,\ast$ &   $3.55$ &    {\it p}  &  $1.18$ &   $12.28$ &   $3.28\pm0.13$ &   $0.29\pm0.04$ &   $\dots$ &   $0.91\pm0.84$ &      &   $\dots$ &      &    f \\ 
  VLAHFF-J041637.88$-$240754.3 &   $0.51$ &    {\it s}  &  $0.99$ &   $10.55$ &   $1.76\pm0.12$ &   $3.29\pm0.22$ &   $4.21\pm0.11$ &   $2.96\pm0.45$ &    $\checkmark$ &   $\dots$ &      &    f \\ 
  VLAHFF-J041641.49$-$240735.3 &   $1.95$ &    {\it p}  &  $1.06$ &   $10.69$ &   $2.86\pm0.13$ &   $5.13\pm0.49$ &   $\dots$ &   $1.48\pm0.57$ &      &   $\dots$ &      &    f \\ 
  VLAHFF-J041641.59$-$240654.0$\,\ast$ &   $2.08$ &    {\it p}  &  $1.23$ &   $10.71$ &   $2.76\pm0.13$ &   $2.47\pm0.38$ &   $\dots$ &   $1.64\pm0.63$ &      &   $\dots$ &      &    f \\ 
  VLAHFF-J071710.16+375000.9 &   $0.18$ &    {\it p}  &  $1.0$ &   $10.24$ &   $0.3\pm0.13$ &   $2.8\pm0.01$ &   $2.77\pm0.07$ &   $0.59\pm0.16$ &      &   $\dots$ &      &    f \\ 
  VLAHFF-J071712.40+374946.8 &   $1.34$ &    {\it p}  &  $1.1$ &   $10.31$ &   $1.88\pm0.15$ &   $2.57\pm7.67$ &   $2.25\pm0.14$ &   $0.0\pm0.76$ &      &   $\dots$ &      &    f \\ 
  VLAHFF-J071715.28+374846.0 &   $1.13$ &    {\it p}  &  $1.09$ &   $10.54$ &   $1.59\pm0.15$ &   $4.52\pm0.16$ &   $4.47\pm0.16$ &   $3.4\pm1.14$ &      &   $\dots$ &      &    f \\ 
  VLAHFF-J071715.71+374801.3 &   $0.42$ &    {\it s}  &  $1.0$ &   $10.49$ &   $0.61\pm0.15$ &   $3.25\pm1.31$ &   $2.78\pm0.22$ &   $1.05\pm0.54$ &      &   $\dots$ &      &    f \\ 
  VLAHFF-J071716.05+375108.5 &   $0.38$ &    {\it p}  &  $1.0$ &   $10.94$ &   $1.32\pm0.13$ &   $5.45\pm0.08$ &   $4.69\pm0.34$ &   $2.83\pm0.39$ &    $\checkmark$ &   $\dots$ &      &    f \\ 
  VLAHFF-J071717.36+374830.4 &   $1.38$ &    {\it p}  &  $1.21$ &   $10.79$ &   $2.04\pm0.13$ &   $6.94\pm0.31$ &   $\dots$ &   $1.23\pm0.4$ &      &   $\dots$ &      &    f \\ 
  VLAHFF-J071717.59+374936.3 &   $0.54$ &    {\it s}  &  $1.0$ &   $10.13$ &   $1.07\pm0.14$ &   $6.73\pm0.41$ &   $5.86\pm0.25$ &   $1.67\pm0.51$ &      &   $\dots$ &      &    f \\ 
  VLAHFF-J071717.94+374918.0 &   $1.57$ &    {\it p}  &  $1.14$ &   $10.75$ &   $2.24\pm0.14$ &   $3.45\pm0.24$ &   $2.28\pm0.2$ &   $2.35\pm0.5$ &    $\checkmark$ &   $\dots$ &      &    f \\ 
  VLAHFF-J071718.25+374840.9 &   $0.68$ &    {\it s}  &  $1.06$ &   $11.02$ &   $1.11\pm0.15$ &   $2.16\pm0.63$ &   $3.58\pm0.07$ &   $0.0\pm1.13$ &      &   $\dots$ &      &    f \\ 
  VLAHFF-J071719.48+374941.4 &   $1.25$ &    {\it p}  &  $1.14$ &   $10.53$ &   $1.96\pm0.14$ &   $2.68\pm0.31$ &   $2.88\pm0.14$ &   $0.93\pm0.67$ &      &   $\dots$ &      &    f \\ 
  VLAHFF-J071719.99+375054.3$\,\ast$ &   $0.85$ &    {\it p}  &  $1.06$ &   $9.44$ &   $1.57\pm0.14$ &   $0.45\pm0.06$ &   $1.98\pm0.61$ &   $1.1\pm0.84$ &      &   $\dots$ &      &    f \\ 
  VLAHFF-J071720.14+374956.8 &   $0.23$ &    {\it p}  &  $1.0$ &   $10.69$ &   $0.45\pm0.19$ &   $\dots$ &   $4.23\pm0.78$ &   $1.87\pm0.49$ &    $\checkmark$ &   $\dots$ &      &    f \\ 
  VLAHFF-J071722.31+375107.8$\,\ast$ &   $1.2$ &    {\it s}  &  $1.1$ &   $10.37$ &   $2.48\pm0.13$ &   $0.38\pm0.06$ &   $1.38\pm0.0$ &   $1.14\pm0.18$ &    $\checkmark$ &   $\dots$ &      &    f \\ 
  VLAHFF-J071723.28+374519.8 &   $0.42$ &    {\it s}  &  $1.0$ &   $10.59$ &   $1.39\pm0.06$ &   $\dots$ &   $3.04\pm0.07$ &   $0.89\pm0.15$ &    $\checkmark$ &   $1.7\pm0.4$ &    $\checkmark$ &    f \\ 
  VLAHFF-J071723.40+374551.8 &   $0.47$ &    {\it p}  &  $1.0$ &   $10.01$ &   $0.81\pm0.14$ &   $\dots$ &   $2.15\pm0.11$ &   $0.97\pm0.4$ &      &   $\dots$ &      &    f \\ 
  VLAHFF-J071724.84+374352.7 &   $0.54$ &    {\it s}  &  $1.0$ &   $10.01$ &   $0.97\pm0.15$ &   $5.01\pm0.09$ &   $4.78\pm0.1$ &   $2.27\pm0.64$ &    $\checkmark$ &   $\dots$ &      &    c \\ 
  VLAHFF-J071724.88+374841.3 &   $1.0$ &    {\it p}  &  $1.08$ &   $10.08$ &   $1.92\pm0.13$ &   $3.39\pm1.16$ &   $3.38\pm0.71$ &   $4.98\pm0.92$ &    $\checkmark$ &   $\dots$ &      &    f \\ 
  VLAHFF-J071724.91+374407.9 &   $0.54$ &    {\it s}  &  $1.0$ &   $10.79$ &   $1.05\pm0.14$ &   $9.33\pm0.86$ &   $8.01\pm0.26$ &   $1.2\pm0.39$ &      &   $\dots$ &      &    c \\ 
  VLAHFF-J071725.18+374354.1 &   $0.73$ &    {\it p}  &  $1.14$ &   $10.44$ &   $1.09\pm0.15$ &   $3.54\pm0.22$ &   $3.05\pm0.03$ &   $1.57\pm0.5$ &      &   $\dots$ &      &    f \\ 
  VLAHFF-J071725.83+375018.9$\,\ast$ &   $2.1$ &    {\it s}  &  $1.05$ &   $10.21$ &   $2.43\pm0.14$ &   $0.38\pm0.11$ &   $1.97\pm0.02$ &   $1.57\pm0.63$ &      &   $\dots$ &      &    f \\ 
  VLAHFF-J071725.85+374446.2 &   $2.93$ &    {\it p}  &  $2.21$ &   $10.36$ &   $2.16\pm0.15$ &   $1.06\pm0.03$ &   $\dots$ &   $1.03\pm0.52$ &      &   $\dots$ &      &    f \\ 
  VLAHFF-J071726.91+374609.2 &   $0.5$ &    {\it s}  &  $1.0$ &   $10.04$ &   $0.75\pm0.15$ &   $2.84\pm0.12$ &   $3.0\pm0.11$ &   $1.36\pm0.63$ &      &   $\dots$ &      &    f \\ 
  VLAHFF-J071727.21+374605.9 &   $0.28$ &    {\it p}  &  $1.0$ &   $9.82$ &   $0.72\pm0.13$ &   $3.94\pm0.02$ &   $3.94\pm0.07$ &   $2.04\pm0.3$ &    $\checkmark$ &   $\dots$ &      &    f \\ 
  VLAHFF-J071727.53+374441.2 &   $0.53$ &    {\it s}  &  $1.0$ &   $10.04$ &   $1.59\pm0.07$ &   $3.38\pm0.29$ &   $3.93\pm0.24$ &   $1.24\pm0.14$ &    $\checkmark$ &   $1.79\pm0.32$ &    $\checkmark$ &    f \\ 
  VLAHFF-J071728.08+374507.4 &   $0.61$ &    {\it p}  &  $1.18$ &   $10.69$ &   $1.31\pm0.13$ &   $5.4\pm0.41$ &   $4.92\pm0.53$ &   $4.29\pm0.94$ &    $\checkmark$ &   $\dots$ &      &    f \\ 
  VLAHFF-J071729.06+374320.0 &   $0.23$ &    {\it s}  &  $1.0$ &   $10.56$ &   $0.04\pm0.15$ &   $4.98\pm2.05$ &   $3.03\pm0.4$ &   $0.54\pm0.35$ &      &   $\dots$ &      &    f \\ 
  VLAHFF-J071729.68+374408.4$\,\ast$ &   $0.55$ &    {\it s}  &  $1.0$ &   $10.77$ &   $0.77\pm0.21$ &   $3.58\pm0.17$ &   $3.77\pm0.03$ &   $1.87\pm0.56$ &    $\checkmark$ &   $1.41\pm0.79$ &      &    c \\ 
  VLAHFF-J071729.75+374524.8 &   $0.56$ &    {\it s}  &  $1.12$ &   $10.79$ &   $1.25\pm0.14$ &   $2.25\pm0.85$ &   $3.05\pm1.77$ &   $2.09\pm0.33$ &    $\checkmark$ &   $\dots$ &      &    f \\ 
  VLAHFF-J071730.39+374617.1 &   $0.91$ &    {\it p}  &  $1.81$ &   $10.62$ &   $1.54\pm0.13$ &   $4.29\pm0.25$ &   $3.56\pm0.16$ &   $3.73\pm0.77$ &    $\checkmark$ &   $\dots$ &      &    f \\ 
  VLAHFF-J071730.65+374443.1 &   $1.01$ &    {\it s}  &  $2.84$ &   $10.43$ &   $0.88\pm0.16$ &   $3.3\pm0.14$ &   $3.33\pm0.05$ &   $1.2\pm0.51$ &      &   $\dots$ &      &    f \\ 
  VLAHFF-J071731.51+374437.5 &   $0.49$ &    {\it s}  &  $1.0$ &   $10.9$ &   $1.12\pm0.14$ &   $3.63\pm0.73$ &   $4.77\pm0.26$ &   $1.54\pm0.27$ &    $\checkmark$ &   $\dots$ &      &    f \\ 
  VLAHFF-J071731.53+374623.8 &   $0.56$ &    {\it p}  &  $1.03$ &   $10.4$ &   $1.16\pm0.14$ &   $5.33\pm0.13$ &   $4.85\pm0.1$ &   $2.89\pm0.66$ &    $\checkmark$ &   $\dots$ &      &    f \\ 
  VLAHFF-J071731.60+374321.3 &   $0.22$ &    {\it p}  &  $1.0$ &   $9.88$ &   $0.87\pm0.04$ &   $1.56\pm0.03$ &   $1.76\pm0.01$ &   $0.94\pm0.07$ &    $\checkmark$ &   $1.03\pm0.15$ &    $\checkmark$ &    f \\ 
  VLAHFF-J071731.75+374333.6 &   $0.53$ &    {\it s}  &  $1.0$ &   $10.29$ &   $1.28\pm0.08$ &   $6.94\pm0.06$ &   $7.47\pm0.17$ &   $2.95\pm0.32$ &    $\checkmark$ &   $2.29\pm0.61$ &    $\checkmark$ &    f \\ 
  VLAHFF-J071731.77+374317.2 &   $0.32$ &    {\it p}  &  $1.0$ &   $9.57$ &   $0.72\pm0.13$ &   $2.26\pm0.02$ &   $2.43\pm0.07$ &   $3.29\pm0.77$ &    $\checkmark$ &   $\dots$ &      &    f \\ 
  VLAHFF-J071732.35+374359.2 &   $0.53$ &    {\it s}  &  $1.0$ &   $9.67$ &   $1.0\pm0.15$ &   $2.27\pm0.13$ &   $2.83\pm0.1$ &   $3.16\pm0.96$ &    $\checkmark$ &   $\dots$ &      &    f \\ 
  VLAHFF-J071732.39+374319.7 &   $0.88$ &    {\it p}  &  $1.25$ &   $10.42$ &   $2.22\pm0.07$ &   $6.21\pm4.36$ &   $4.39\pm0.07$ &   $0.68\pm0.17$ &    $\checkmark$ &   $0.0\pm0.87$ &      &    f \\ 
  VLAHFF-J071733.14+374543.2 &   $0.91$ &    {\it s}  &  $2.11$ &   $9.89$ &   $1.38\pm0.14$ &   $1.36\pm0.07$ &   $1.27\pm0.03$ &   $2.26\pm0.47$ &    $\checkmark$ &   $\dots$ &      &    f \\ 
  VLAHFF-J071734.46+374432.2 &   $1.14$ &    {\it s}  &  $5.84$ &   $9.42$ &   $1.56\pm0.12$ &   $1.13\pm0.35$ &   $2.02\pm0.14$ &   $2.69\pm0.37$ &    $\checkmark$ &   $\dots$ &      &    f \\ 
  VLAHFF-J071735.13+374552.7$\,\ast$ &   $0.55$ &    {\it s}  &  $1.01$ &   $10.06$ &   $0.59\pm0.18$ &   $3.75\pm0.14$ &   $4.27\pm0.23$ &   $0.0\pm0.34$ &      &   $\dots$ &      &    c \\ 
  VLAHFF-J071735.22+374541.7$\,\ast$ &   $1.69$ &    {\it s}  &  $3.61$ &   $10.87$ &   $1.73\pm0.14$ &   $1.38\pm0.21$ &   $1.89\pm0.07$ &   $0.0\pm0.24$ &      &   $\dots$ &      &    f \\ 
  VLAHFF-J071735.30+374447.3 &   $0.18$ &    {\it s}  &  $1.0$ &   $10.08$ &   $0.32\pm0.13$ &   $1.93\pm0.21$ &   $1.7\pm0.1$ &   $1.99\pm0.33$ &    $\checkmark$ &   $\dots$ &      &    f \\ 
  VLAHFF-J071735.65+374517.1 &   $0.54$ &    {\it s}  &  $1.0$ &   $12.0$ &   $1.43\pm0.09$ &   $5.17\pm2.01$ &   $5.86\pm0.17$ &   $0.0\pm0.05$ &      &   $0.74\pm0.62$ &      &    c \\ 
  VLAHFF-J071736.66+374506.4 &   $1.13$ &    {\it s}  &  $6.45$ &   $9.48$ &   $1.01\pm0.15$ &   $1.8\pm0.04$ &   $\dots$ &   $1.33\pm0.35$ &    $\checkmark$ &   $\dots$ &      &    f \\ 
  VLAHFF-J071737.75+374530.0 &   $0.55$ &    {\it s}  &  $1.0$ &   $9.84$ &   $1.02\pm0.14$ &   $3.33\pm0.15$ &   $3.06\pm0.17$ &   $1.19\pm0.34$ &      &   $\dots$ &      &    c \\ 
  VLAHFF-J071738.33+374600.0 &   $0.71$ &    {\it p}  &  $1.17$ &   $10.84$ &   $1.08\pm0.15$ &   $3.45\pm0.98$ &   $2.21\pm0.06$ &   $0.27\pm1.21$ &      &   $1.23\pm0.66$ &      &    f \\ 
  VLAHFF-J071740.24+374306.5 &   $1.93$ &    {\it p}  &  $1.49$ &   $10.21$ &   $1.88\pm0.15$ &   $0.94\pm0.09$ &   $\dots$ &   $0.0\pm0.65$ &      &   $\dots$ &      &    f \\ 
  VLAHFF-J071740.30+374445.4 &   $0.55$ &    {\it s}  &  $1.01$ &   $9.08$ &   $1.16\pm0.15$ &   $1.79\pm0.06$ &   $2.19\pm0.1$ &   $2.39\pm0.52$ &    $\checkmark$ &   $\dots$ &      &    c \\ 
  VLAHFF-J071740.55+374506.4$\,\ast$ &   $1.97$ &    {\it p}  &  $2.18$ &   $10.48$ &   $1.83\pm0.15$ &   $\dots$ &   $\dots$ &   $0.0\pm0.51$ &      &   $\dots$ &      &    f \\ 
  VLAHFF-J071741.28+374452.1 &   $0.57$ &    {\it s}  &  $1.03$ &   $10.33$ &   $0.96\pm0.14$ &   $3.69\pm0.07$ &   $3.35\pm0.02$ &   $1.11\pm0.53$ &      &   $\dots$ &      &    f \\ 
  VLAHFF-J071741.56+374555.5 &   $0.26$ &    {\it p}  &  $1.0$ &   $10.84$ &   $1.07\pm0.1$ &   $7.32\pm1.2$ &   $4.2\pm0.34$ &   $2.9\pm0.35$ &    $\checkmark$ &   $1.13\pm0.4$ &      &    f \\ 
  VLAHFF-J071742.24+374336.0 &   $0.54$ &    {\it s}  &  $1.0$ &   $11.01$ &   $1.31\pm0.15$ &   $\dots$ &   $3.92\pm0.07$ &   $1.49\pm0.25$ &    $\checkmark$ &   $2.35\pm0.8$ &      &    c \\ 
  VLAHFF-J071743.17+374651.1 &   $0.56$ &    {\it s}  &  $1.01$ &   $10.37$ &   $1.08\pm0.15$ &   $\dots$ &   $2.99\pm0.02$ &   $2.73\pm0.84$ &    $\checkmark$ &   $\dots$ &      &    f \\ 
  VLAHFF-J114929.44+222314.3 &   $0.33$ &    {\it p}  &  $1.0$ &   $10.1$ &   $0.69\pm0.06$ &   $3.79\pm0.02$ &   $3.84\pm0.17$ &   $1.92\pm0.26$ &    $\checkmark$ &   $2.28\pm0.49$ &    $\checkmark$ &    f \\ 
  VLAHFF-J114930.67+222427.7 &   $1.49$ &    {\it s}  &  $1.73$ &   $11.01$ &   $2.1\pm0.13$ &   $5.06\pm1.51$ &   $4.64\pm0.2$ &   $3.63\pm0.73$ &    $\checkmark$ &   $\dots$ &      &    f \\ 
  VLAHFF-J114930.80+222327.0 &   $0.37$ &    {\it s}  &  $1.0$ &   $10.44$ &   $1.25\pm0.04$ &   $3.77\pm0.04$ &   $3.54\pm0.33$ &   $0.72\pm0.04$ &    $\checkmark$ &   $0.8\pm0.08$ &    $\checkmark$ &    f \\ 
  VLAHFF-J114930.83+222253.9 &   $0.41$ &    {\it p}  &  $1.0$ &   $11.01$ &   $1.45\pm0.03$ &   $8.7\pm0.08$ &   $6.49\pm1.18$ &   $1.26\pm0.06$ &    $\checkmark$ &   $1.09\pm0.09$ &    $\checkmark$ &    f \\ 
  VLAHFF-J114931.31+222252.1 &   $0.52$ &    {\it p}  &  $1.0$ &   $10.66$ &   $-0.02\pm0.07$ &   $7.13\pm0.14$ &   $6.65\pm0.48$ &   $0.85\pm0.29$ &      &   $0.82\pm0.51$ &      &    f \\ 
  VLAHFF-J114932.01+221754.4 &   $2.16$ &    {\it p}  &  $0.99$ &   $10.58$ &   $2.51\pm0.14$ &   $2.77\pm0.24$ &   $\dots$ &   $0.0\pm0.55$ &      &   $\dots$ &      &    f \\ 
  VLAHFF-J114932.03+222439.3 &   $1.28$ &    {\it s}  &  $2.11$ &   $10.19$ &   $2.12\pm0.19$ &   $3.81\pm0.28$ &   $4.09\pm0.34$ &   $2.13\pm0.4$ &    $\checkmark$ &   $1.52\pm0.53$ &      &    f \\ 
  VLAHFF-J114933.01+222313.3 &   $0.56$ &    {\it s}  &  $1.01$ &   $9.48$ &   $0.71\pm0.08$ &   $2.58\pm0.47$ &   $3.46\pm0.41$ &   $1.27\pm0.27$ &    $\checkmark$ &   $4.03\pm0.83$ &    $\checkmark$ &    f \\ 
  VLAHFF-J114933.14+222430.2 &   $0.55$ &    {\it s}  &  $1.08$ &   $10.68$ &   $0.63\pm0.17$ &   $2.79\pm0.6$ &   $3.55\pm0.09$ &   $0.0\pm0.2$ &      &   $0.71\pm0.65$ &      &    c \\ 
  VLAHFF-J114933.57+222321.7 &   $0.97$ &    {\it s}  &  $1.46$ &   $10.03$ &   $1.19\pm0.13$ &   $1.7\pm0.16$ &   $2.37\pm1.08$ &   $0.95\pm0.24$ &    $\checkmark$ &   $1.78\pm0.28$ &    $\checkmark$ &    f \\ 
  VLAHFF-J114933.77+222533.2 &   $1.39$ &    {\it p}  &  $1.34$ &   $10.75$ &   $1.43\pm0.17$ &   $2.94\pm0.65$ &   $\dots$ &   $1.02\pm0.69$ &      &   $\dots$ &      &    f \\ 
  VLAHFF-J114933.88+222226.9$\,\ast$ &   $2.61$ &    {\it p}  &  $1.22$ &   $11.66$ &   $3.47\pm0.05$ &   $0.89\pm0.4$ &   $1.66\pm0.01$ &   $1.52\pm0.04$ &    $\checkmark$ &   $1.66\pm0.12$ &    $\checkmark$ &    f \\ 
  VLAHFF-J114934.46+222438.5 &   $0.75$ &    {\it s}  &  $2.16$ &   $9.77$ &   $0.72\pm0.13$ &   $2.44\pm0.15$ &   $2.84\pm0.23$ &   $1.05\pm0.26$ &    $\checkmark$ &   $1.31\pm0.51$ &      &    f \\ 
  VLAHFF-J114934.65+222320.8 &   $0.96$ &    {\it s}  &  $1.53$ &   $10.43$ &   $1.14\pm0.15$ &   $1.03\pm2.5$ &   $2.18\pm0.04$ &   $1.27\pm0.26$ &    $\checkmark$ &   $2.64\pm0.71$ &    $\checkmark$ &    f \\ 
  VLAHFF-J114935.47+222231.9 &   $0.43$ &    {\it p}  &  $1.0$ &   $8.89$ &   $0.73\pm0.15$ &   $3.77\pm0.11$ &   $3.4\pm0.12$ &   $1.32\pm0.4$ &    $\checkmark$ &   $\dots$ &      &    f \\ 
  VLAHFF-J114936.09+222424.4 &   $1.64$ &    {\it p}  &  $3.13$ &   $10.71$ &   $1.67\pm0.15$ &   $4.43\pm1.47$ &   $3.5\pm0.22$ &   $1.58\pm0.35$ &    $\checkmark$ &   $\dots$ &      &    f \\ 
  VLAHFF-J114936.83+222253.6 &   $1.41$ &    {\it s}  &  $1.6$ &   $10.32$ &   $2.21\pm0.11$ &   $2.51\pm0.03$ &   $2.57\pm0.02$ &   $2.96\pm0.49$ &    $\checkmark$ &   $2.36\pm0.57$ &    $\checkmark$ &    f \\ 
  VLAHFF-J114936.85+222346.9 &   $0.53$ &    {\it s}  &  $1.0$ &   $10.99$ &   $-0.27\pm0.19$ &   $2.96\pm0.42$ &   $3.34\pm0.03$ &   $0.0\pm0.93$ &      &   $\dots$ &      &    c \\ 
  VLAHFF-J114936.98+222542.0 &   $0.5$ &    {\it p}  &  $1.0$ &   $9.42$ &   $1.13\pm0.19$ &   $4.75\pm0.12$ &   $6.59\pm0.22$ &   $2.03\pm0.44$ &    $\checkmark$ &   $\dots$ &      &    f \\ 
  VLAHFF-J114937.62+222536.0 &   $0.52$ &    {\it p}  &  $1.0$ &   $9.52$ &   $1.19\pm0.08$ &   $2.56\pm0.02$ &   $2.54\pm0.0$ &   $1.62\pm0.23$ &    $\checkmark$ &   $1.58\pm0.5$ &      &    f \\ 
  VLAHFF-J114937.99+222427.7 &   $0.53$ &    {\it s}  &  $1.0$ &   $10.07$ &   $0.78\pm0.11$ &   $7.29\pm1.54$ &   $4.04\pm0.43$ &   $0.98\pm0.21$ &    $\checkmark$ &   $1.46\pm0.36$ &    $\checkmark$ &    c \\ 
  VLAHFF-J114938.11+222411.4 &   $1.03$ &    {\it s}  &  $1.68$ &   $10.4$ &   $1.65\pm0.13$ &   $4.22\pm0.29$ &   $3.27\pm0.17$ &   $3.4\pm0.76$ &    $\checkmark$ &   $\dots$ &      &    f \\ 
  VLAHFF-J114938.62+222145.3 &   $2.59$ &    {\it p}  &  $1.2$ &   $10.62$ &   $2.36\pm0.15$ &   $2.02\pm0.15$ &   $\dots$ &   $1.32\pm0.46$ &      &   $\dots$ &      &    f \\ 
  VLAHFF-J114938.92+222259.8 &   $0.54$ &    {\it s}  &  $1.0$ &   $11.28$ &   $1.69\pm0.02$ &   $3.23\pm0.71$ &   $3.9\pm0.21$ &   $0.31\pm0.15$ &      &   $0.27\pm0.09$ &      &    c \\ 
  VLAHFF-J114939.34+222227.3 &   $0.84$ &    {\it p}  &  $1.18$ &   $9.73$ &   $1.08\pm0.16$ &   $4.41\pm0.15$ &   $4.0\pm0.24$ &   $1.16\pm0.63$ &      &   $\dots$ &      &    f \\ 
  VLAHFF-J114939.73+221856.6 &   $0.54$ &    {\it p}  &  $1.0$ &   $8.74$ &   $1.09\pm0.14$ &   $1.87\pm0.38$ &   $1.2\pm0.17$ &   $0.44\pm0.34$ &      &   $\dots$ &      &    f \\ 
  VLAHFF-J114939.79+222503.2 &   $0.92$ &    {\it p}  &  $1.19$ &   $10.59$ &   $1.77\pm0.13$ &   $1.41\pm0.27$ &   $2.9\pm0.21$ &   $0.95\pm0.16$ &    $\checkmark$ &   $0.88\pm0.36$ &      &    f \\ 
  VLAHFF-J114940.15+222233.3 &   $0.92$ &    {\it p}  &  $1.28$ &   $10.25$ &   $1.32\pm0.13$ &   $5.09\pm0.33$ &   $4.36\pm0.25$ &   $1.6\pm0.27$ &    $\checkmark$ &   $1.3\pm0.39$ &      &    f \\ 
  VLAHFF-J114940.57+222415.6 &   $0.77$ &    {\it s}  &  $1.18$ &   $10.24$ &   $1.33\pm0.14$ &   $5.23\pm0.53$ &   $4.48\pm0.39$ &   $2.05\pm0.48$ &    $\checkmark$ &   $\dots$ &      &    f \\ 
  VLAHFF-J114940.86+222307.4 &   $0.93$ &    {\it s}  &  $1.31$ &   $10.23$ &   $1.17\pm0.15$ &   $3.57\pm0.31$ &   $2.87\pm0.22$ &   $1.67\pm0.59$ &      &   $\dots$ &      &    f \\ 
  VLAHFF-J114942.38+222339.5 &   $1.48$ &    {\it p}  &  $1.36$ &   $9.73$ &   $2.13\pm0.27$ &   $3.93\pm0.16$ &   $\dots$ &   $1.64\pm0.32$ &    $\checkmark$ &   $1.35\pm0.49$ &      &    f \\ 
  VLAHFF-J114943.08+221745.0 &   $1.04$ &    {\it p}  &  $1.04$ &   $11.22$ &   $2.58\pm0.11$ &   $7.86\pm5.0$ &   $6.57\pm0.23$ &   $0.49\pm0.15$ &      &   $0.9\pm0.59$ &      &    f \\ 
  VLAHFF-J114943.72+222412.8 &   $0.48$ &    {\it p}  &  $1.0$ &   $9.35$ &   $1.68\pm0.11$ &   $3.33\pm0.01$ &   $3.67\pm0.11$ &   $2.38\pm0.33$ &    $\checkmark$ &   $1.34\pm0.43$ &      &    f \\ 
  VLAHFF-J114944.33+222408.6 &   $0.93$ &    {\it p}  &  $1.09$ &   $10.07$ &   $1.44\pm0.25$ &   $3.49\pm0.29$ &   $4.14\pm0.01$ &   $1.95\pm0.51$ &    $\checkmark$ &   $1.53\pm0.69$ &      &    f \\ 
  VLAHFF-J114945.47+221700.4 &   $4.43$ &    {\it p}  &  $1.03$ &   $11.55$ &   $3.8\pm0.13$ &   $3.33\pm0.25$ &   $\dots$ &   $1.24\pm0.12$ &    $\checkmark$ &   $\dots$ &      &    f \\ 
  VLAHFF-J114945.73+222419.0 &   $0.58$ &    {\it p}  &  $1.01$ &   $10.08$ &   $1.17\pm0.14$ &   $3.87\pm0.78$ &   $2.83\pm0.04$ &   $2.56\pm0.61$ &    $\checkmark$ &   $\dots$ &      &    f \\ 
  VLAHFF-J114948.03+221907.2 &   $2.16$ &    {\it p}  &  $1.24$ &   $9.8$ &   $2.22\pm0.16$ &   $1.98\pm0.1$ &   $\dots$ &   $0.86\pm0.59$ &      &   $\dots$ &      &    f \\ 
  VLAHFF-J114948.12+221830.6 &   $0.75$ &    {\it p}  &  $1.08$ &   $10.47$ &   $1.36\pm0.14$ &   $0.66\pm0.01$ &   $1.67\pm0.02$ &   $1.39\pm0.42$ &      &   $\dots$ &      &    f \\ 
  VLAHFF-J114949.00+221803.2 &   $2.77$ &    {\it p}  &  $1.39$ &   $10.62$ &   $2.42\pm0.16$ &   $\dots$ &   $\dots$ &   $0.0\pm0.78$ &      &   $\dots$ &      &    f \\ 
    \enddata 
	\tablenotetext{a}{Type of redshift: {\it s}$-$spectroscopic/{\it p}$-$photometric.}\vspace{-0.1cm}
	\tablenotetext{b}{Total SFR: $\rm SFR_{3\,GHz}+SFR_{FUV}$.}\vspace{-0.1cm}
	\tablenotetext{c}{Flag: reliably resolved source at 3\,GHz.}\vspace{-0.1cm}
	\tablenotetext{d}{Flag: reliably resolved source at 6\,GHz.}\vspace{-0.1cm}
	\tablenotetext{c}{Environment of galaxies: f$-$field/c$-$cluster.}\vspace{-0.1cm}
	\tablenotetext{\ast}{  Radio source with reported X-ray counterpart (see \S \ref{subsec:size-starformation} for details)}   
	\label{table_sllsizes}
\end{deluxetable*}
\end{longrotatetable}

\onecolumngrid
\section{The impact of AGN candidates in the size -- SFR, stellar mass, and redshift relations.}
\begin{figure*}[h!]
	\begin{centering}
		\includegraphics[width=18cm]{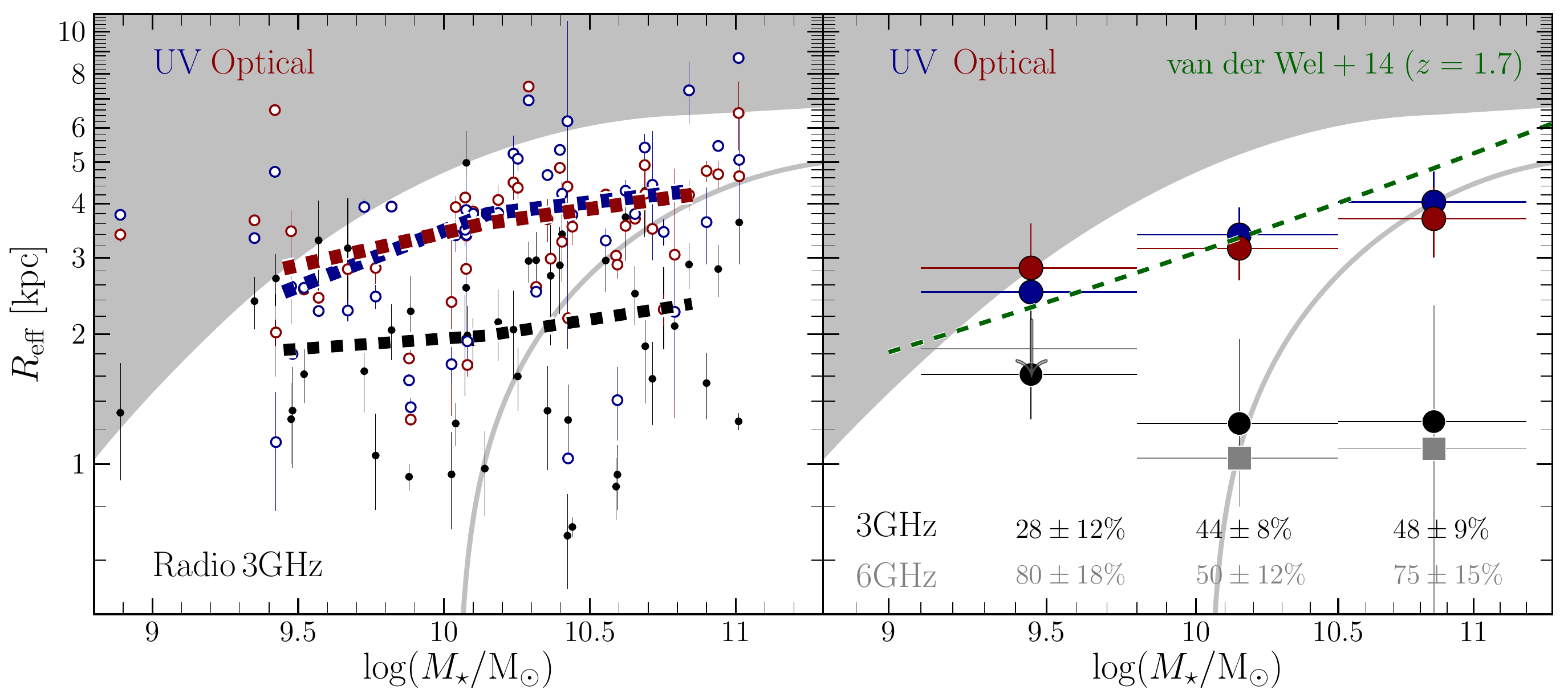}	
		\caption{
		 \color{black} {\it Left panel--} The radio/UV/optical effective radius of galaxies that are reliably resolved at 3\,GHz as a function of their stellar mass. The dashed lines show the median effective radius per stellar mass bin (0.75\,dex width).  {\it Right panel--} The median radio/UV/optical effective radius of all galaxies in the sample (i.e., sources that are reliably and unreliably resolved at 3\,GHz) per stellar mass bin. Contrary to the results presented in Figure\,\ref{fig:size-mass}, here we exclude the radio sources that have an X-ray counterpart, which are likely to host an AGN. A complete description of this figure can be found in the caption of Figure\,\ref{fig:size-mass}.}
		\label{fig:size-mass_noagn}
	\end{centering}
\end{figure*}

\begin{figure*}[h!]
	\begin{centering}
		\includegraphics[width=18cm]{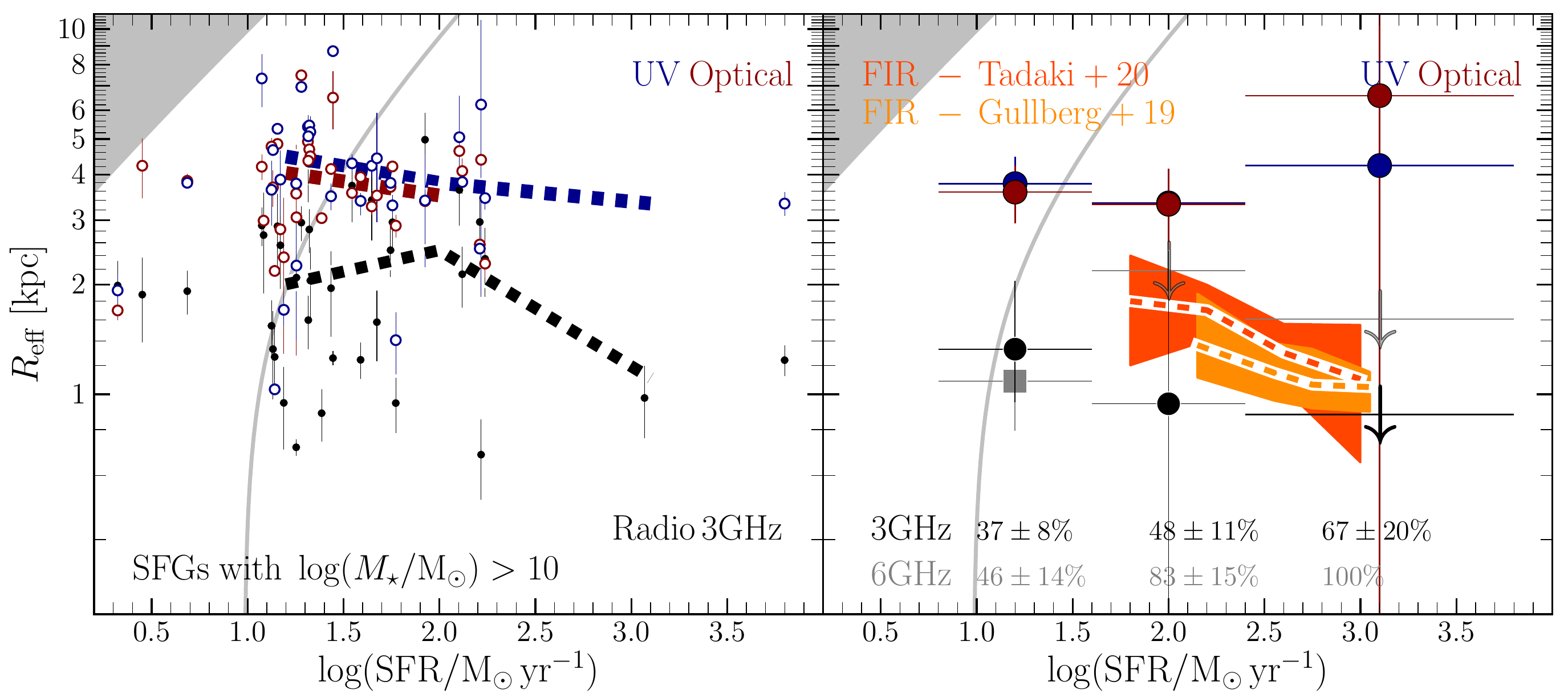}	
		\caption{  \color{black}  {\it Left panel--} The radio/UV/optical effective radius of $\log(M_\star/\rm M_\odot)>10$ SFGs that are reliably resolved at 3\,GHz as a function of their SFR. The dashed lines show the median effective radius per SFR bin ($1-1.5$\,dex width).  {\it Right panel--} The median radio/UV/optical effective radius  of all galaxies in the sample (i.e., sources that are reliably and unreliably resolved at 3\,GHz) per SFR bin. Contrary to the results presented in Figure\,\ref{fig:size-sfr}, here we exclude the radio sources that have an X-ray counterpart, which are likely to host an AGN. A complete description of this figure can be found in the caption of Figure\,\ref{fig:size-sfr}.     }
		\label{fig:size-sfr_noagn}
	\end{centering}
\end{figure*}

\begin{figure*}
	\begin{centering}
		\includegraphics[width=18cm]{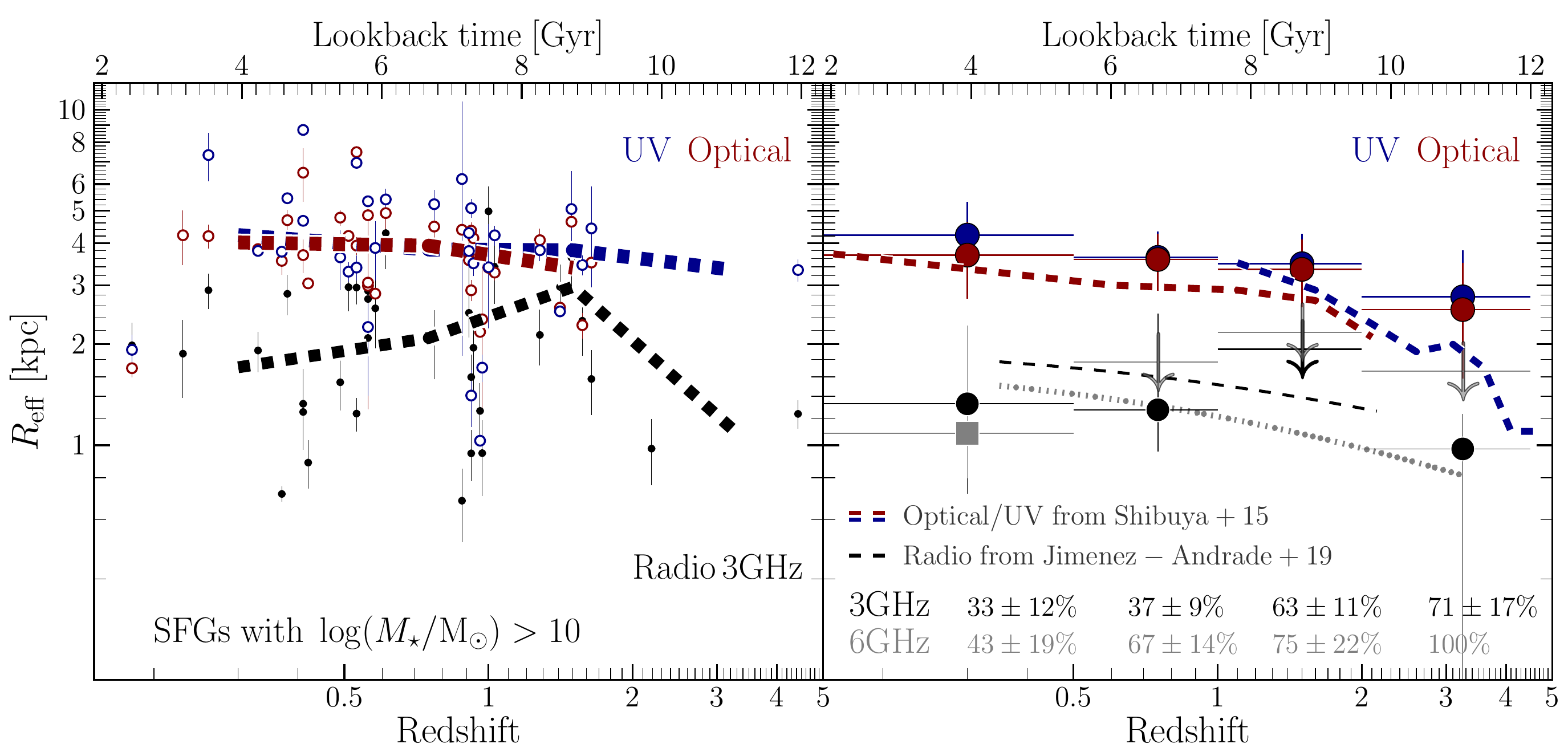}	
		\caption{
		 \color{black} {\it Left panel--} Redshift evolution of the radio/UV/optical effective radius  of SFGs with $\log(M_\star/\rm M_\odot)>10$  that are reliably resolved at 3\,GHz. The dashed lines show the median effective radius per redshift bin ($\approx 2$\,Gyr width).   {\it Right panel--} The median radio/UV/optical effective radius of all galaxies in the sample (i.e., sources that are reliably and unreliably resolved at 3\,GHz) per redshift bin. The values at the bottom indicate the fraction of unreliably resolved galaxies at 3 and 6\,GHz, respectively.   Contrary to the results presented in Figure\,\ref{fig:size-sfr}, here we exclude the radio sources that have an X-ray counterpart, which are likely to host an AGN.  The dotted blue line is a  fit to the 3\,GHz data points presented here, leading to $R_{\rm eff}=(1.7\pm 0.4) (1+z)^{-0.5\pm 0.4}$\,kpc.	A more complete description of this figure can be found in the caption of Figure\,\ref{fig:size-redshift}. }   
		\label{fig:size-redshift_noagn}
	\end{centering}
\end{figure*}

{   \color{black} Out of 113 galaxies in our sample, 12 have an X-ray counterpart with a luminosity $L_{\rm X}>10^{42}\rm \,erg\,s^{-1}$. Thus, it is likely that the radio emission from these  galaxies have some potential contribution from an AGN (\S \ref{subsub:agn_fraction}). By removing the 12 AGN candidates from the analysis, we find that  the radio size -- SFR, stellar mass, and redshift relations (Figure\,\ref{fig:size-mass_noagn}, \ref{fig:size-sfr_noagn}, \ref{fig:size-redshift_noagn}) are consistent with the trends derived from the full galaxy sample (Figure\,\ref{fig:size-mass}, \ref{fig:size-sfr}, \ref{fig:size-redshift}). The only exception is the median UV/optical size of galaxies in our highest SFR bin, which is strongly affected by small-number statistics  after removing the AGN candidates from the sample.}

\clearpage


\bibliographystyle{aasjournal}
\bibliography{jimenezandrade+20_vlahff.bib} 



\end{document}